\pdfoutput=1
\documentclass[onecolumn]{aastex}
\usepackage{graphicx}
\usepackage{amsmath}
\usepackage{hyperref}
\usepackage{threeparttable}
\DeclareGraphicsExtensions{.pdf,.eps}
\usepackage[utf8]{inputenc}
\usepackage{cleveref}
\usepackage{natbib}
\crefname{section}{\S}{\S \S}

\shorttitle{MC$^2$: A deeper look at ZwCl 2341.1+0000}
\shortauthors{Benson et al.}

\begin{document}
\title{MC$^2$: A deeper look at ZwCl~2341.1+0000 with Bayesian galaxy clustering and weak lensing analyses}

\author{B. Benson, D. M. Wittman, N. Golovich}
\affil{Physics Department, University of California, Davis, CA, 95616}

\author{M. James Jee\altaffilmark{1}}
\affil{Department of Astronomy and Center for Galaxy Evolution Research, Yonsei University, 50 Yonsei-ro, Seoul 03772, Korea}

\author{R. J. van Weeren}
\affil{Harvard Smithsonian Center for Astrophysics, 60 Garden Street, Cambridge, MA 02138}

\and 

\author{W. A. Dawson}
\affil{Lawrence Livermore National Laboratory, Livermore, CA, 94550}

\altaffiltext{1}{Physics Department, University of California, Davis, CA, 95616}

\received{22 March 2017}

\accepted{12 April 2017}

\begin{abstract}
ZwCl 2341.1+0000, a merging galaxy cluster with disturbed X-ray morphology and widely separated ($\sim$3 Mpc) double radio relics, was thought to be an extremely massive ($10-30 \times 10^{14} M_\odot$) and complex system with little known about its merger history.  We present JVLA 2-4 GHz observations of the cluster, along with new spectroscopy from our Keck/DEIMOS survey, and apply Gaussian Mixture Modeling to the three-dimensional distribution of 227 confirmed cluster galaxies.  After adopting the Bayesian Information Criterion to avoid overfitting, which we discover can bias total dynamical mass estimates high, we find that a three-substructure model with a total dynamical mass estimate of $9.39 \pm 0.81 \times 10^{14} M_\odot$ is favored.  We also present deep Subaru imaging and perform the first weak lensing analysis on this system, obtaining a weak lensing mass estimate of $5.57 \pm 2.47 \times 10^{14} M_\odot$.  This is a more robust estimate because it does not depend on the dynamical state of the system, which is disturbed due to the merger.  Our results indicate that ZwCl 2341.1+0000 is a multiple merger system comprised of at least three substructures, with the main merger that produced the radio relics occurring near to the plane of the sky, and a younger merger in the North occurring closer to the line of sight.  Dynamical modeling of the main merger reproduces observed quantities (relic positions and polarizations, subcluster separation and radial velocity difference), if the merger axis angle of $\sim$10$^{+34}_{-6}$ degrees and the collision speed at pericenter is $\sim$1900$^{+300}_{-200}$ km/s.
\end{abstract}


\section{Introduction}

Merging galaxy clusters are of interest in cosmology for their richness in astrophysical processes, including shocks, enhanced X-ray and nonthermal radio emissions, and their potential to probe the nature of dark matter.  Each merging cluster system observed, however, provides only a single snapshot in the long merger history, and we must infer many of the details necessary to reconstruct these merger scenarios.  Piecing together the information from previously observed systems, a general understanding of the merger process has emerged.  This begins when two galaxy clusters experience a mutual gravitational attraction, causing them to fall towards each other.  As the clusters cross at the pericenter of the merger, the ionized gas in their intra-cluster mediums will experience a momentum exchange due to Coulomb interaction.  In this process some of the gas is often stripped from the parent clusters, leaving that gas lagging behind the other components.  Galaxies, in contrast, are essentially collisionless and exit the core crossing with very little momentum lost.  From observations of numerous other merging cluster systems, we know that the dark matter will also exit the core crossing with little to no offset between it and the galaxies \citep{Randall07, Dawson12, Ng15}.

When the two clusters interact during the core crossing, a shock is launched through the gas in the system that can be detectable in the X-ray and/or radio \citep{Ensslin98, Markevitch02}.  In some cases, extended and polarized radio emission features, called radio relics, are generated at the shock fronts preceding the clusters as they move away from the pericenter.  Radio relics are generally observable in mergers taking place in or near the plane of the sky, which makes the relics useful for identifying systems where a separation between cluster components (i.e. gas, galaxies, and dark matter) is most visible.  The polarization of a radio relic can also help to constrain the angle the merger axis makes with the plane of the sky \citep{Ng15, Ensslin98}, while the orientation of the relic can be used to constrain the azimuthal angle of the merger axis \citep{vanWeeren11}.  In some galaxy cluster mergers a double relic (one on either side of the system) is observed, which can help to more precisely pin down the merger axis.   

\begin{figure}
\plotone{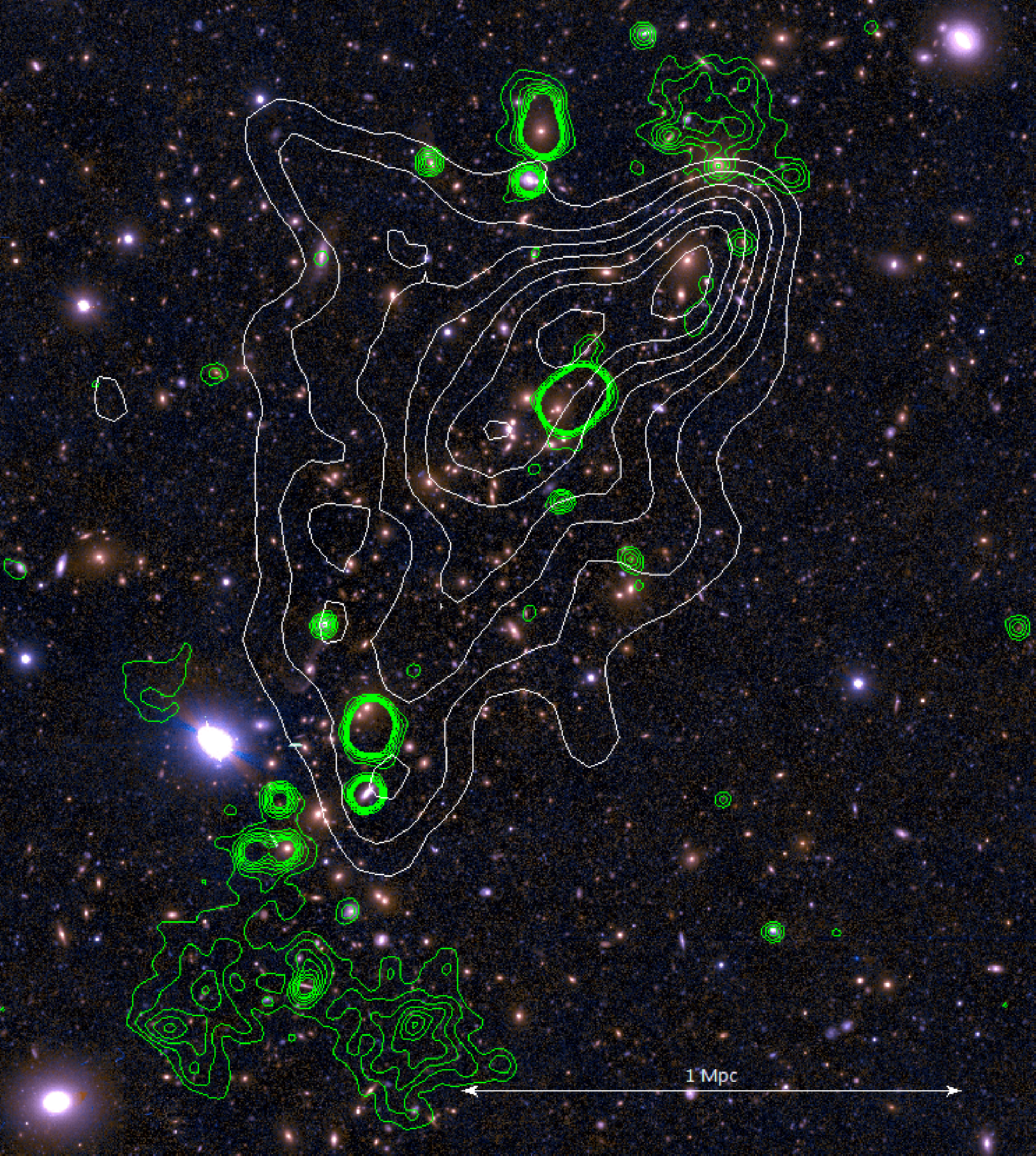}
\caption{Subaru color image of ZwCl 2341.1+0000 with Chandra X-ray contours (white) and GMRT radio contours (green) overlayed \citep{vanWeeren09}.  The diffuse radio emissions to the North-West and South-East indicate the two radio relics.  These relics bracket the elongated and disturbed X-ray emissions, which is typical of a galaxy cluster merger.\label{fig1}}
\end{figure}

The Merging Cluster Collaboration (MCC) has selected 28 clusters possessing radio relics from the \cite{Feretti12} review for a lensing and spectroscopic survey.  ZwCl 2341.1+0000 is one of these selected clusters and sits at a redshift of 0.27.  Figure \ref{fig1} shows the $\sim$3 Mpc separated radio relics \citep{vanWeeren09} bracketing a disturbed X-ray emitting gas cloud \citep{Bagchi02} for this system.

Previous works on ZwCl 2341.1+0000 include an optical analysis and dynamical mass estimate using data from the Sloan Digital Sky Survey (SDSS) and spectroscopy on 101 galaxies at the cluster redshift (\cite{Boschin13}, hereafter referred to as B13).  B13 concluded that the system is comprised of likely three to six subclusters with a total dynamical mass estimate in the range of $10-30 \times10^{14} h^{-1}_{70} M_\odot$.  The large mass, as well as its complex nature as either a multiple merger between three or more galaxy clusters, or as a node in the cosmic web with multiple in-falling filaments, make ZwCl 2341.1+0000 a very interesting system.  However, dynamical mass estimates tend to be biased high in merging galaxy clusters due to the disturbed nature of the galaxy velocities, which are no longer in equilibrium with the gravitational potential to which they are bound.  \cite{Pinkney96} found in simulations that for a 3:1 mass ratio merger, the velocity dispersion could be boosted to $\sim$2 times the equilibrium value, which could lead to over-estimating the dynamical mass by a factor of four.

One of our primary goals in studying ZwCl 2341.1+0000 is to obtain the first weak lensing mass estimate of the system, which will better constrain the cluster mass because it is not dependent upon the dynamical state of the clusters, and thus is not biased in an active merger.  We also seek to pin down the number of substructures needed to describe ZwCl 2341.1+0000 by more than doubling the number of spectroscopically confirmed cluster members and applying our Gaussian Mixture Model clustering analysis.  These results are then used for the dynamical modeling of ZwCl 2341.1+0000, as described in \cite{Dawson13}, to help estimate some of the merger properties.

The structure of this paper is as follows.  In \S 2 we describe our data.  We discuss the optical analyses, including the galaxy cluster substructures of the merger, and the dynamical mass estimate in \S 3.  Weak lensing analysis is discussed in \S 4, and we conclude in \S 5.  We indicate all uncertainties at the 68 per cent confidence level.  Throughout this paper we adopt a flat, $\Lambda$CDM cosmology, with $H_0 = 70$ km s$^{-1}$Mpc$^{-1}$, $h = 0.7$, $\Omega_m = 0.3$, and $\Omega_\Lambda = 0.7$.  In this adopted cosmology the length scale is 4.14 kpc/arcsec at the cluster redshift of z = 0.27.

\section[]{Observations and Data}

\subsection{Subaru}
We observed ZwCl 2341.1+0000 using the Subaru Prime Focus Camera (SuprimeCam) with the 8.2m Subaru telescope at the National Astronomical Observatory of Japan on Mauna Kea (P.I. D. Wittman).  We carried out the observations over two nights (2013 February 23-24), taking four 180.0 second exposures in the g$^{\prime}$ band and eight 360.0 second exposures in the r$^{\prime}$ band.  The telescope was rotated between each exposure to distribute the bleeding trails and diffraction spikes from bright stars, which were later removed by median stacking the images.  This maximized the number of source galaxies usable for weak lensing analysis.  The median seeing for both bands was $\sim0.53$".  

Reduction of the optical data was performed in the same manner as \cite{Jee15, Jee16}, with the SDFRED \citep{Ouchi04, Yagi02}, SCAMP \citep{Bertin06}, and SWARP \citep{Bertin02} packages.  The first was used to subtract over-scan and bias, make flats, correct for geometric distortion, and mask regions affected by bad pixels and auto guide probe, and the second to remove residual distortion, refine the astrometrics, and correct for photometric calibration differences between exposures.  The last was used to create a large $\sim 40.5' \times 42.5'$ mosaic image based on the distortion and flux calibration solutions output by SCAMP.

\subsection{DEIMOS}
We conducted a spectroscopic survey of ZwCl 2341.1+0000 with the DEIMOS instrument on the Keck II 10m telescope over two separate observing runs, with two observations performed on 2013 July 14 and one performed on 2013 September 05.  Our objective with these observations was to target a maximum number of the brightest cluster member galaxies.  For this, we used the SDSS DR9 \citep{Ahn12} publicly available imaging with photometric redshift estimations to select galaxies in the bright end of the identified red sequence with photo-z estimates near the cluster redshift.  The SDSS catalog was used for this because our own Subaru images were not available at the time of the planning.  

All observations were taken with 1" wide slits and the 1200 line mm$^{-1}$ grating.  This resulted in a pixel scale of 0.33 \AA pixel$^{-1}$ and a resolution of $\sim$ 1 \AA.  Our goal was to observe the H$\beta$, [O III], Mg I (b), Fe I, Na I (D), [O I], H$\alpha$, and/or the [N II] doublet for each galaxy in order to secure accurate redshifts for as many of our targets as possible.  Reduction of the spectroscopic data was performed with a modified DEEP2 version of the spec2d package \citep{Newman13}.  We observed a total of three slit masks with approximately 120 slits per mask.  For each mask we took three 900 second exposures.  We adopted the DEEP 2 quality rating system \citep{Newman13} and only accepted redshifts rated as secure or very secure (3 or 4).  This yielded a total of 301 usable redshifts with an average uncertainty of $3.13 \times10^{-5}$ (cluster rest frame velocity uncertainty of 6.33 km/s) from our survey.

\subsection{JVLA}
ZwCl 2341.1+0000 was observed with the Karl G. Jansky Very Large Array (JVLA) in the $\sim2-4$ GHz S-band in D-array and C-array, and an overview of the observations is given in Table \ref{tab:jvlaobs}.  The data were reduced with {\tt CASA} \citep{McMullin07} version 4.5.  Two different pointing centers were observed, with one centered on the northern relic and the other on the southern relic.  Below we briefly summarize the data reduction, but for more details the reader is referred to \cite{vanWeeren16}.

As a first step in the data reduction we removed radio frequency interference (RFI) by employing the {\tt AOFlagger} \citep{Offringa10}, and the `tfcrop' mode in the task {\tt flagdata} in {\tt CASA}.  Data affected by antenna shadowing were also flagged.  Delay, bandpass, cross-hand delay, gain, polarization leakage and angles calibration solutions were obtained using observations of calibrator sources.  These calibration tables were then applied to the target field data.  The datasets from each pointing and array configuration were first imaged separately, then the calibration of the individual datasets were further refined via the process of self-calibration.  These datasets from the different array configurations were then combined and again imaged.  One additional round of self-calibration was carried out to precisely align the datasets from the different configurations.  During the self-calibration imaging, we used Briggs weighting {\tt robust=0} and employed W-projection \citep{Cornwell08, Cornwell05}.  The spectral index was taken into account during the deconvolution \citep{Rau11} and clean masks were used throughout the process.  These masks were made with the {\tt PyBDSM} source detection package \citep{Mohan15}.

After the calibration, we made separate images for each of the two pointings.  We combined these pointings in the image plane to make a single image of the cluster, correcting for the primary beam attenuation in the process, producing both low- and high-resolution continuum images.  The low-resolution image was made with natural weighting, which resulted in a beam of $32\arcsec \times 24\arcsec$ and a noise level of $\sigma_{\rm{rms}}= 15 ~\mu$Jy beam$^{-1}$.  The high-resolution image was made with uniform weighting, resulting in a beam of $6.3\arcsec \times 5.6\arcsec$ and $\sigma_{\rm{rms}}=13 ~\mu$Jy beam$^{-1}$.

\begin{figure}
\includegraphics[height=6.8in, width=5.8in, angle=90.0]{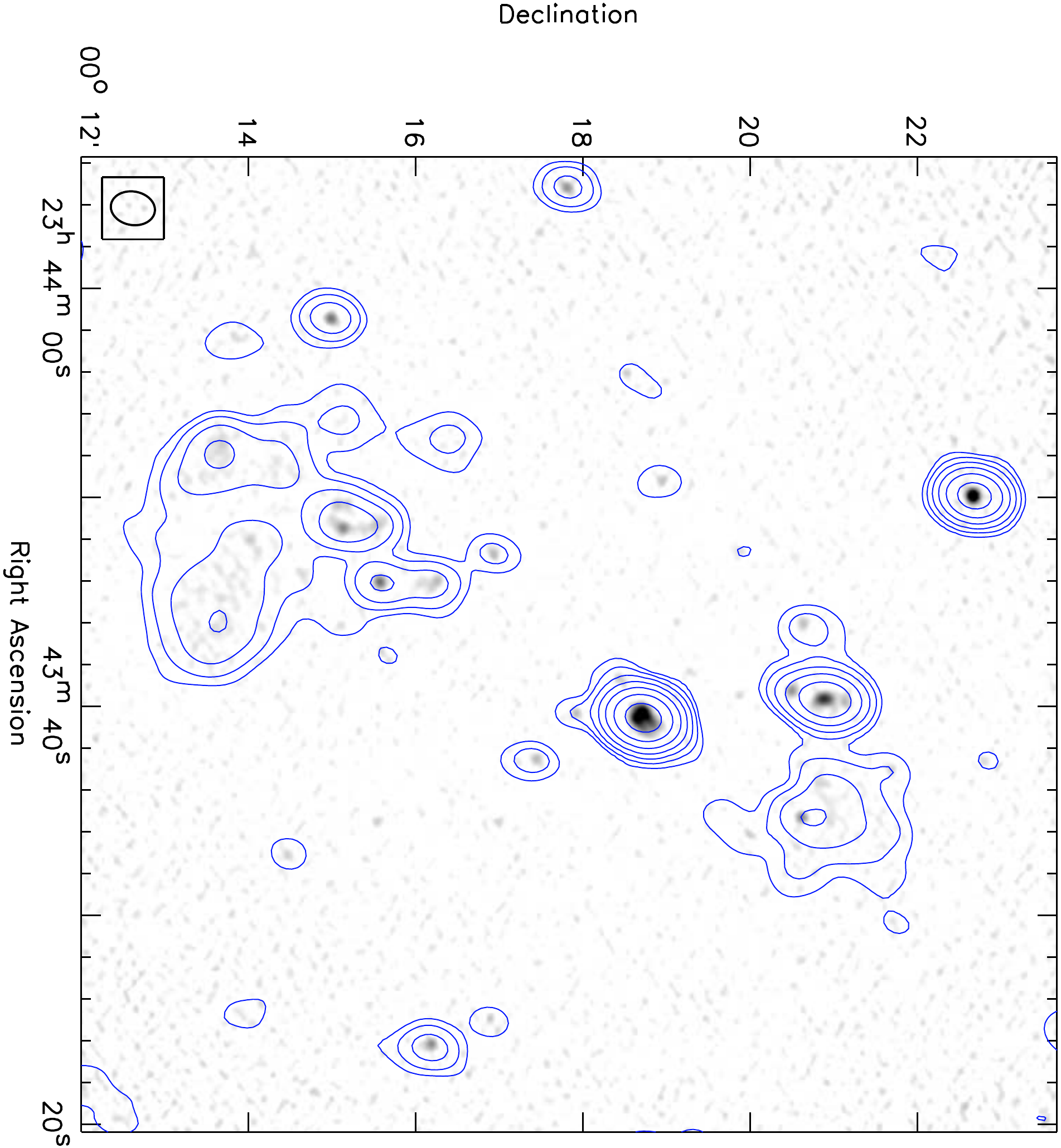}
\caption{S-band 2--4~GHz continuum image of ZwCl~2341.1+0000 in greyscale. Blue contours are from the low-resolution continuum image and overlaid at levels of $[1,2,4,8, \ldots] \times 4\sigma_{\rm{rms}}$. The beam size is indicated in the bottom left corner.\label{fig2}}
\end{figure}

The high-resolution image, overlaid with the low-resolution contours, is displayed in Figure \ref{fig2}.  In this image, the most prominent source is a central ``tailed'' radio galaxy \citep{vanWeeren09}, which the high-resolution S-band image suggests could be classified as a narrow angle tailed (NAT) radio galaxy, with the tail pointing NNE.  Only hints of the two relics are seen in the high-resolution image due to their low surface brightness.  The morphologies of the relics in the low-resolution image are similar to previous observations at lower frequencies \citep{Bagchi02, vanWeeren09, Giovannini10}, and no evidence for additional extended emission was found in our S-band image.

\begin{figure} 
\includegraphics[height=6.8in, width=5.8in, angle=90.0]{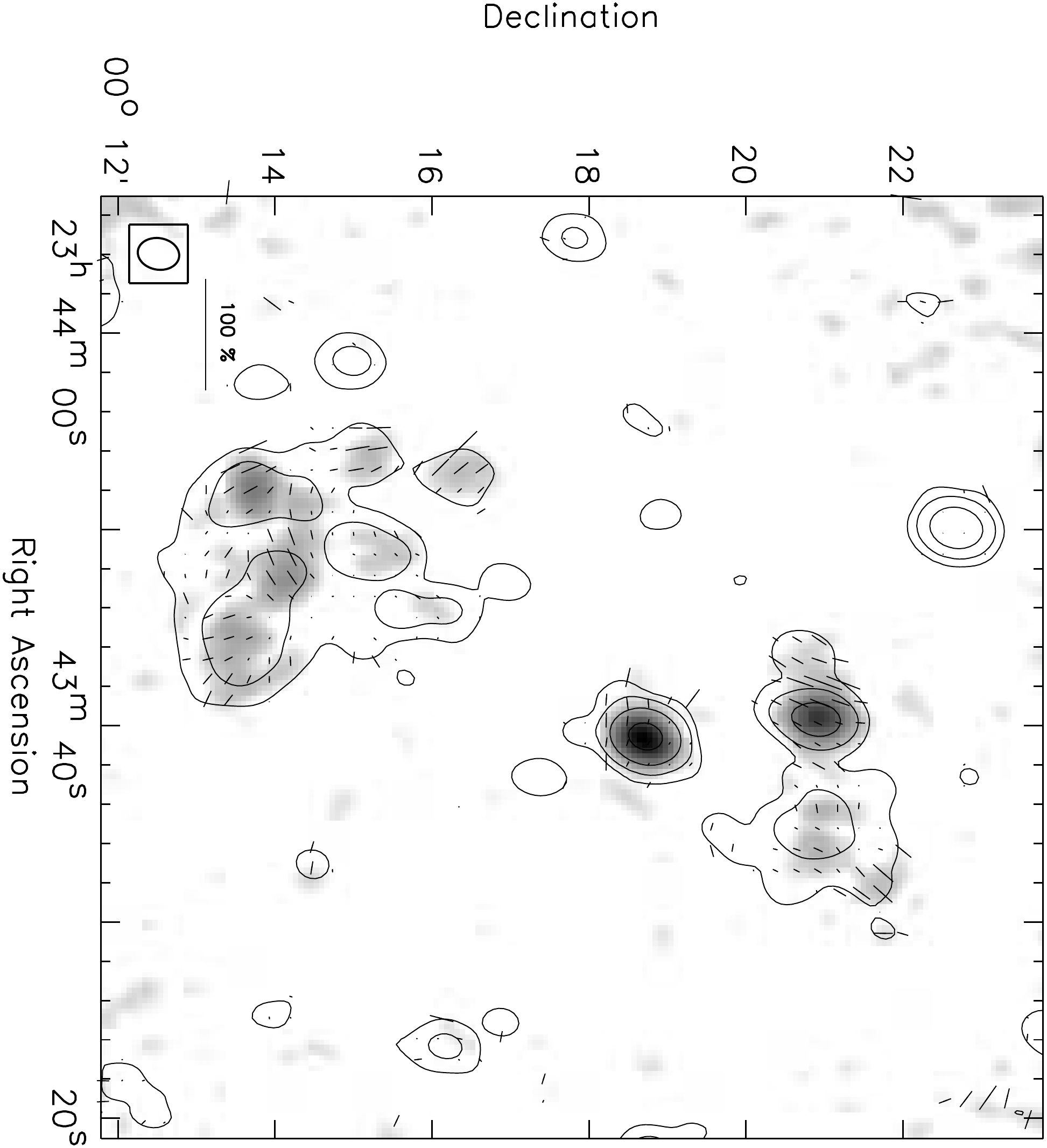} 
\caption{S-band 2--4~GHz polarization image of ZwCl~2341.1+0000.1+0000. The greyscale image displays the total polarized intensity ($P$) image ($P = \sqrt{Q^2 + U^2}$). The vectors depict the electric field direction. The length of the vectors is proportional to the polarization fraction ($\frac{P}{I}$). A reference vector for 100\% polarization is shown in the bottom left corner.
Black contours are the continuum (Stokes I) image. Contours are drawn at levels of $[1,4,16, 64, \ldots] \times 4\sigma_{\rm{rms}}$. The beam size is indicated in the bottom left corner.\label{fig3}}
\end{figure}

Given the low amount of Galactic Faraday Rotation in the direction of ZwCl~2341.1+0000 \citep[a Rotation Measure (RM) of $\approx -4$ rad m$^{-2}$][]{Taylor09}, we produced Stokes Q and U images using the entire available bandwidth.  To make these images, we employed the same weighting scheme as for the low-resolution continuum image.  The polarization electric field vector map and polarized intensity map are shown in Figure \ref{fig3}.  Our polarization image revealed polarized flux from both relics, albeit the S/N was relatively low due to the low-surface brightness of the relics.  Polarized flux of the relics was previously detected at 1.4~GHz by \cite{Giovannini10}.  The orientation of the polarization vectors around the northern relic (in particular its northern edge) and the bright AGN next to it  \citep{vanWeeren09} were very similar to what was reported by \cite{Giovannini10}, which was also consistent with the low amount of Galactic Faraday Rotation and the peripheral location of the relic emission.  Due to our higher spatial resolution, we observed more details for the southern relic.  The polarization fraction and angles varied considerably across this relic, with measurements as high as 20--30\% in a few locations.  However, the integrated polarization fraction for both relics was much lower, measuring only 5\% and  8\%, respectively, for the northern and southern relics.  While most relics in the literature have higher polarization fractions reported, these are usually not emission weighted average values, and it therefore remains to be determined if these values are anomalously low.  An example of this can be seen in the relic of Abell 3411-3412, where the maximum polarization fraction is $\sim$40\%, but the emission weighted average value is only 13\% \citep{vanWeeren17}.

\begin{table}
\begin{center}
\caption{JVLA Observations}
\begin{tabular}{lllll}
&S-band D-array & S-band C-array & \\
\hline
\hline
Observation dates &    Oct 19, 2015 &  Oct 9, 2014 \\
Frequencies coverage (GHz)      &  2--4 & 2--4 & \\
On source time per pointing  (hr)   & 1.5 & 0.5    \\
Channel width (MHz) & 2 &  2  \\
Integration time (s)     & 5 & 5 \\
LAS$^{a}$ (arcsec)           &  490 & 490 \\
\hline
\end{tabular}
\label{tab:jvlaobs}
\end{center}
$^{a}$ Largest angular scale that can be recovered by these observations. \\
\end{table}

We extracted the flux densities in polygon regions enclosing the two relics from the low-resolution continuum image.  Emissions from compact sources embedded in the relics were subtracted by measuring the flux densities of the compact sources in the high-resolution image.  The uncertainties on the relics’ integrated flux densities were computed from the rms map noise, scaling with the source area and including an absolute flux calibration uncertainty of 5\%.  We find a 3~GHz integrated flux density of $1.73 \pm 0.11$~mJy and $6.40\pm 0.34$~mJy for the North and South relics, respectively.  Using the reported 610~MHz flux densities from the GMRT \citep{vanWeeren09}, we compute integrated spectral indices of $-1.31\pm0.14$ for the North relic, and $-1.20\pm 0.18$ for the South relic.  These values are consistent with those reported by \cite{Giovannini10} and are typical values for radio relics \citep[e.g.,][]{vanWeeren09-2,Kale12,Stroe13,Hindson14}.

\subsection{Additional Data Used}
In addition to the Deimos targeting, we also used the SDSS DR9 \citep{Ahn12} publicly available imaging in our analysis of ZwCl 2341.1+0000.  We used the positions and photometric redshift estimations, as well as dereddened model magnitudes in g and r band for all galaxies with r band magnitude less than 22 in our field of interest.

Our spectroscopic data were collected before B13 was published, leading to an overlap of 71 galaxies in our targeted objects, of which 59 were cluster galaxies.  Of these 59 galaxies, the mean difference between our redshifts and those obtained by B13 was $\Delta z = 0.00025 \pm 0.000078$.  Given how small the differences were, we use our values for all overlapping objects, and the B13 values for the remaining 55 unique redshifts measured in their survey, as well as 6 additional redshifts from the public NED\footnote{The NASA/IPAC Extragalactic Database (NED)
is operated by the Jet Propulsion Laboratory, California Institute of Technology,
under contract with the National Aeronautics and Space Administration.} database.

\section[]{Galaxy Cluster Member Selection}
For the next stages in the analysis, we determined which galaxies were members of the ZwCl 2341.1+0000 cluster.  Cluster membership was determined in two ways, the first of which was based on the spectroscopic redshifts.  This catalog had the advantage of being a pure sample of galaxies in our target volume of space, but it suffered from a lack of completeness.  To complement this, we compiled a photometric catalog by using our spectroscopically confirmed cluster members to identify a red sequence for the cluster.  We then applied red sequence cuts to the entire SDSS catalog previously mentioned and our own Subaru catalog.  This sample was photometrically complete, but also included some foreground and background galaxies.  In each of these catalogs we limited our galaxies to objects within $\sim$8' of the Subaru field center, RA = $23^h43^m37.44^s$, Dec = $0^h16^m23.44^s$.  This removed the strong vignetting in the outer regions of the Subaru SuprimeCam, as well as outliers that are likely too far separated to be cluster members (8' equates to $\sim$2 Mpc at redshift 0.27).

\subsection{Spectroscopic Redshift Selection}
\begin{figure}  
\plotone{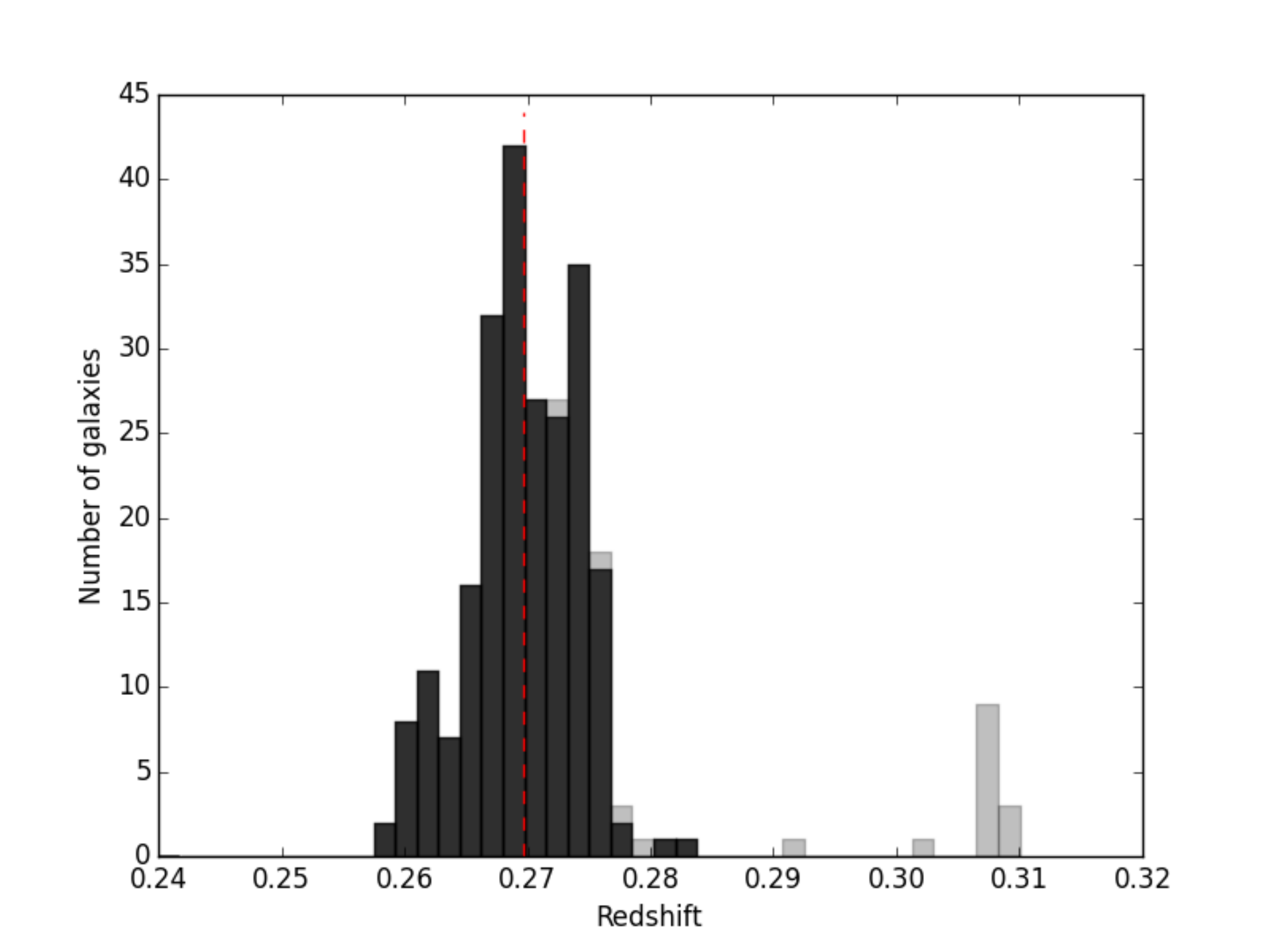} 
\caption{Histogram of all redshifts in our spectroscopic catalog with cluster galaxies shown in black, non-cluster galaxies in grey, and the mean redshift of the cluster galaxies marked with a dashed red line.  Non-cluster galaxies within the cluster redshift window were too far separated (more than 2 Mpc) from the cluster center to be considered as cluster members.\label{fig4}} 
\end{figure}

\begin{figure} 
\plotone{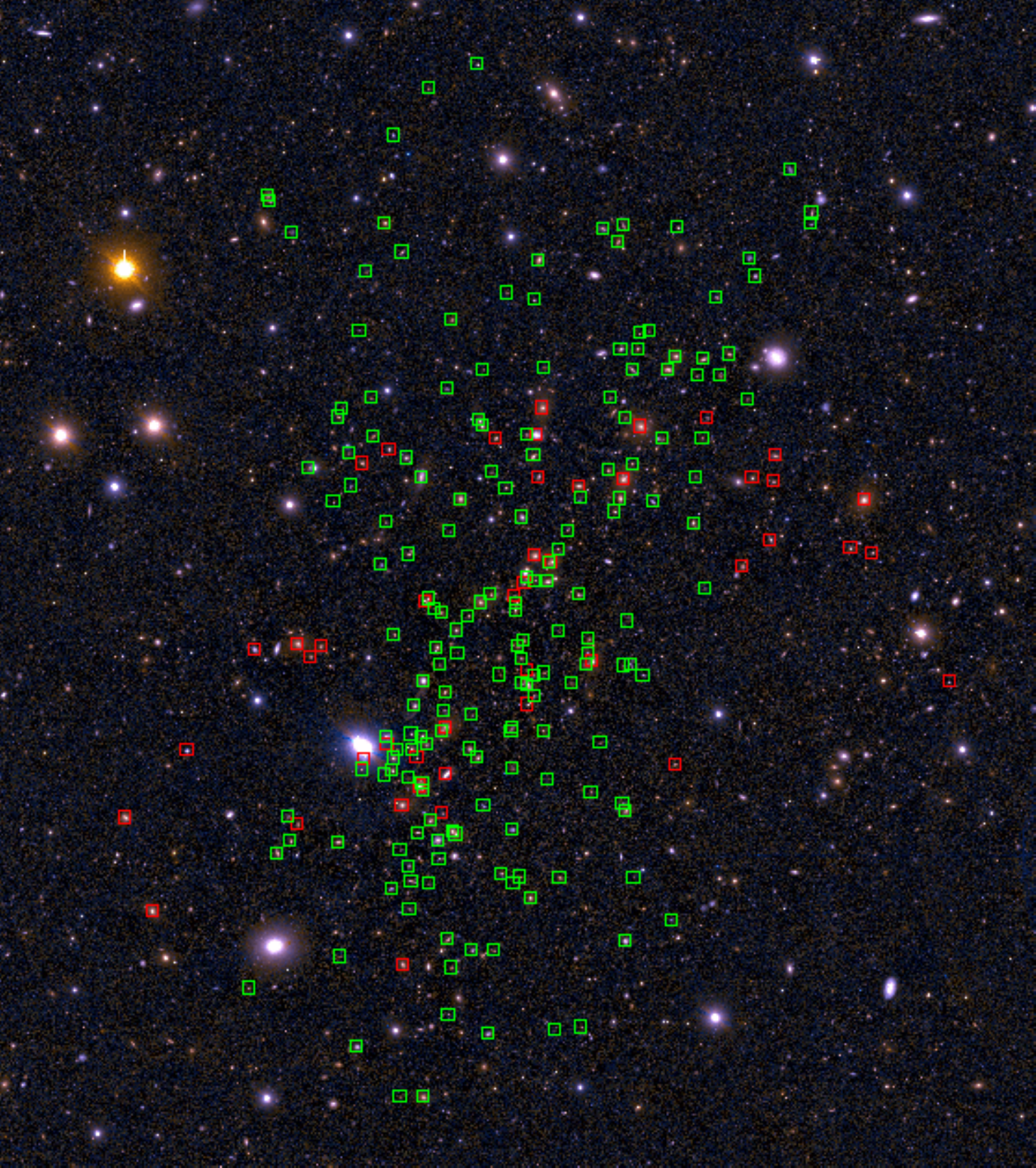} 
\caption{Subaru image of ZwCl 2341.1+0000 with green boxes marking the spectroscopically confirmed cluster members obtained from our Deimos observations and red boxes marking the cluster members obtained from B13 and NED.\label{fig5}}
\end{figure}

Figure \ref{fig4} presents a histogram of the spectroscopic redshifts in our combined spectroscopic catalog near the cluster redshift, which peaks at 0.27.  We considered all objects within 3$\sigma_v$ ($\sim$ 3000 km/s) of the peak to be cluster members.  This yielded a selection criteria of $0.256\leq z \leq 0.283$, and resulted in 227 cluster members for our spectroscopic catalog.  Figure \ref{fig5} presents an image of the system with all spectroscopically confirmed cluster galaxies marked, and Table \ref{tab:Velocities} lists their positions and velocities.

\begin{deluxetable}{ccrrrrrrrrcrl}
\tabletypesize{\scriptsize}
\tablecaption{List of positions and redshifts for confirmed cluster galaxies\label{tab:Velocities}}
\tablewidth{0pt}
\tablehead{
\colhead{RAh} & \colhead{RAm} & \colhead{RAs} & \colhead{DE-} & \colhead{DEd} & \colhead{DEm} & \colhead{DEs} & \colhead{z} & \colhead{$\sigma_z$}
}
\startdata
23 & 43 & 47.29 & + & 0 & 10 & 53.18 & 0.271744 & 0.000017 \\
23 & 43 & 48.65 & + & 0 & 10 & 53.33 & 0.271749 & 0.000075 \\
23 & 43 & 51.15 & + & 0 & 11 & 37.07 & 0.271268 & 0.000005 \\
23 & 43 & 43.55 & + & 0 & 11 & 48.05 & 0.268486 & 0.000005 \\
23 & 43 & 39.65 & + & 0 & 11 & 51.43 & 0.272691 & 0.000006 \\
23 & 43 & 38.11 & + & 0 & 11 & 54.17 & 0.274128 & 0.000073 \\
23 & 43 & 45.83 & + & 0 & 12 & 5.26 & 0.271259 & 0.000044 \\
23 & 43 & 57.43 & + & 0 & 12 & 28.19 & 0.273506 & 0.000017 \\
23 & 43 & 45.67 & + & 0 & 12 & 45.83 & 0.266173 & 0.000023 \\
23 & 43 & 48.49 & + & 0 & 12 & 48.70 & 0.265057 & 0.000032 \\
\enddata

\tablecomments{Table \ref{tab:Velocities} is published in its entirety in the 
electronic edition of the {\it Astrophysical Journal}.  A portion is 
shown here for guidance regarding its form and content.}
\end{deluxetable}

\subsection{Red Sequence/Photometric Selection}
\begin{figure} 
\plotone{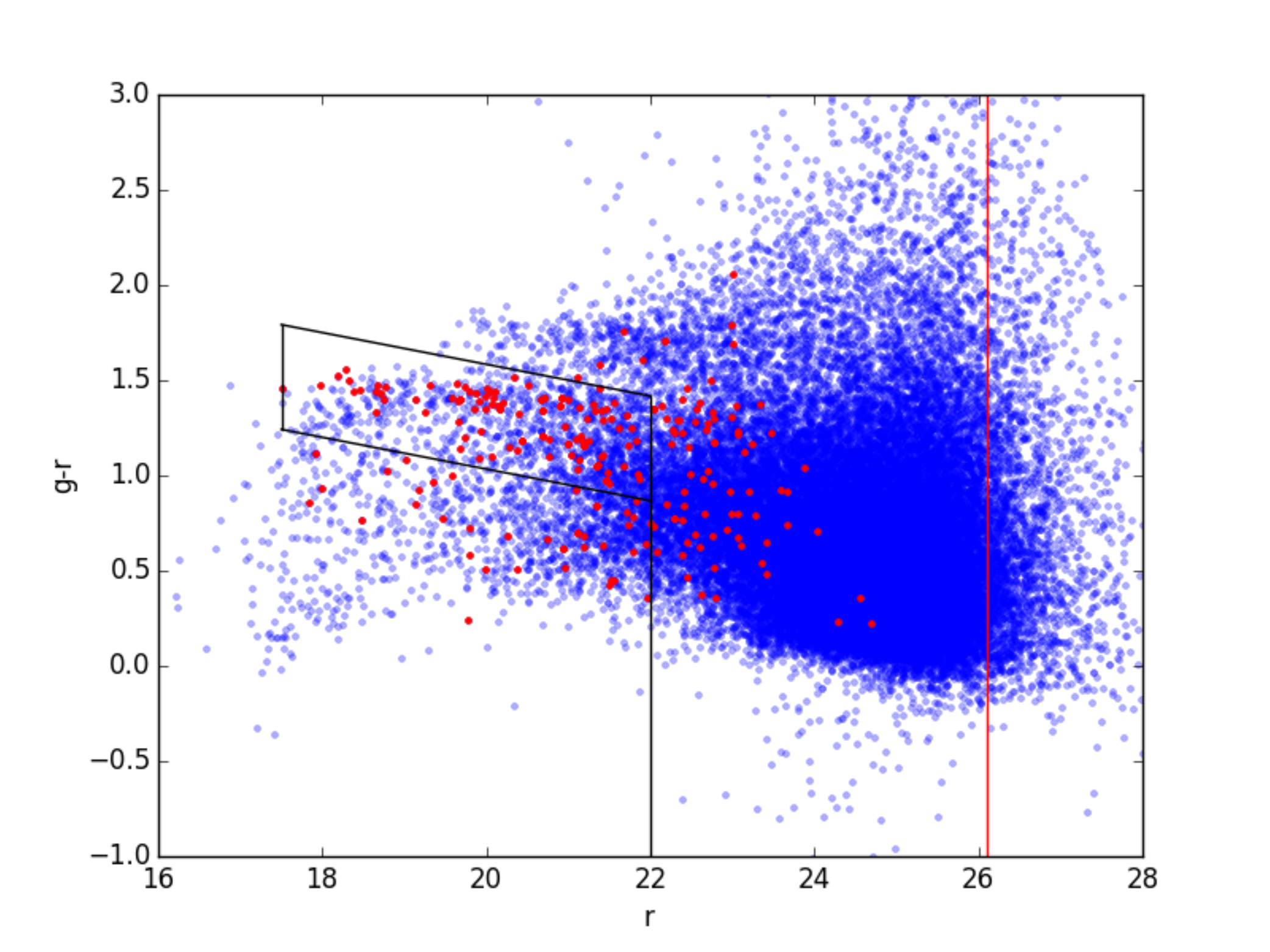} 
\caption{Color vs. magnitude plot for galaxies in the full Subaru catalog (blue) and galaxies from the spectroscopic catalog (red) with the red sequence for ZwCl 2341.1+0000 clearly visible in both.  There also appears to be a second red sequence slightly above the one we are interested in, but the galaxies that make up this collection are not spatially correlated, and do not form a background cluster.  The area inside the black box indicates the color-magnitude selection used on the SDSS data to select red sequence cluster members for the photometric catalog.  Galaxies above and to the right of the black lines, and to the left of the red line, were selected for the weak lensing catalog.\label{fig6}} 
\end{figure}

Figure \ref{fig6} presents a color-magnitude plot of g$^{\prime}$-r$^{\prime}$ vs. r$^{\prime}$ for galaxies in our Subaru SuprimeCam observations (blue), along with the confirmed cluster members from our spectroscopic catalog (red).  This diagram shows a distinct red sequence for ZwCl 2341.1+0000.  This red sequence, along with the spectroscopically confirmed cluster galaxies, informed a set of color and magnitude cuts (shown as a box in Figure \ref{fig6}) similar to those used by B13 on their SDSS catalog to select cluster members.  We also chose to use the SDSS catalog over our Subaru catalog for this selection after we determined that SDSS goes deep enough to include the most useful part of the red sequence (r$^{\prime}$ magnitude less than 22) for the purpose of selecting potential cluster members.  The SDSS photometric redshift estimations also provided much better foreground/background discrimination than our two-band Subaru data.  Our selection criteria for red sequence cluster members in the photometric catalog are $18 \leq r^{\prime} \leq 22$, color satisfying $-0.0833r^{\prime} + 2.7 \leq g^{\prime}-r^{\prime} \leq -0.0833r^{\prime} +3.25$ and $0.19 \leq z_{phot} \leq 0.35 $.

\section[]{Optical Analysis}
In this section we attempt to identify the substructures that make up the larger system to help determine how many clusters are participating in the merger.  We then calculate a dynamical mass estimate for each of these substructures.

\subsection{Galaxy Cluster Substructures}
\begin{figure} 
\plottwo{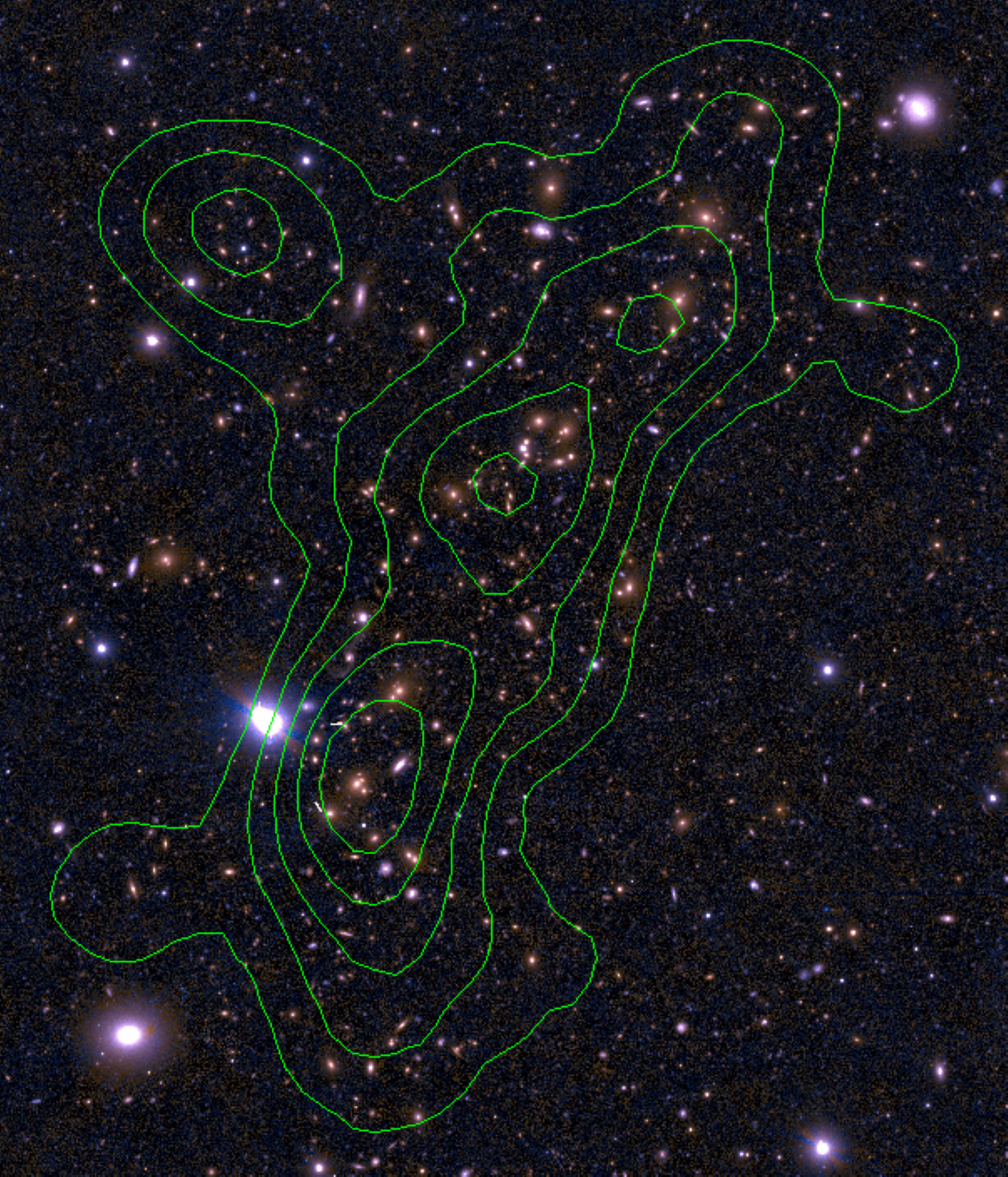}{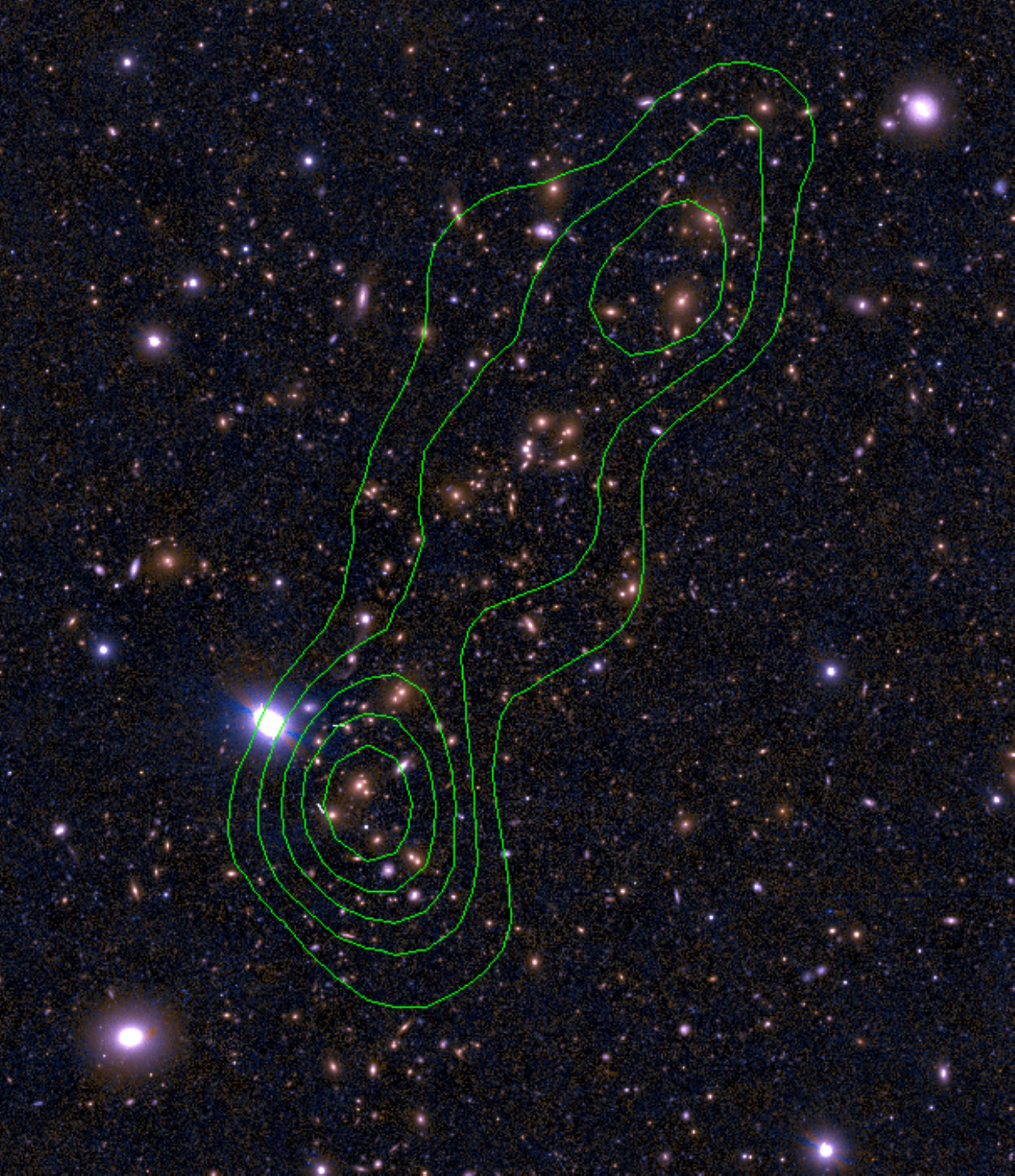}
\caption{(Left) Subaru image of ZwCl 2341.1+0000 with overlayed contours indicating the smoothed projected number-density of red sequence selected galaxies from the photometric catalog. (Right) Subaru image of ZwCl 2341.1+0000 with overlayed iso-luminosity contours from the same red sequence selection of the photometric catalog.  We only used the red sequence galaxies for these plots because their brightness is a more accurate tracer of cluster mass than the bluer disk galaxies. The majority of red light from cluster galaxies is concentrated in the South-East region of the $\sim$3 Mpc elongated body of the system.\label{fig7}} 
\end{figure}

Our first step in identifying the substructures in ZwCl 2341.1+0000 was to perform a bootstrap galaxy centroiding \citep{Dawson12}.  This began with producing a smoothed projected number density plot from the red sequence galaxies in the photometric catalog, shown in Figure \ref{fig7} (Left).  The number density plot was then used to identify overdense regions of galaxies in the system, of which there could be many in the elongated structure of ZwCl 2341.1+0000 depending on the amount of smoothing used.  We smoothed the number density field with a Gaussian kernel of $\sigma$= 80 arcsec.

The first contour was then chosen to contain $\sim$50 galaxies/Mpc$^2$ , which was 3$\sigma$ above the mean background.  The mean galaxy density increases by $\sim$15 galaxies/Mpc$^2$ for each successive contour.  We also created a luminosity density map (Figure \ref{fig7} Right) from the same catalog after we confirmed that the brightest galaxies in the catalog were the brightest cluster galaxies from our spectroscopic survey.  The same smoothing was used on this luminosity map as discussed above.  Following the assumption that brighter galaxies are associated with areas of deeper gravitational potential, a luminosity density map should be a better indicator of where the mass in the system was concentrated.  This has been confirmed by mock observations of Illustris simulations (Ng et al., in preparation).  

From these two maps we can see that both the most overdense region of galaxies and the highest concentration of light are in the South-East end of the system.  However, the light is much more concentrated in the South-East region than the number density, as shown by how much tighter the contours are in the luminosity map.  There are also several disconnected groups of galaxies in the number density map that did not have enough collective brightness to appear above the background in the luminosity map.  Because we are looking at two dimensional projections, it is not clear whether these are significant substructures in ZwCl 2341.1+0000.  We opted to use these maps only to inform three-dimensional approaches to determine the substructures in ZwCl 2341.1+0000, which can better judge if galaxies are spatially correlated, rather than use the results from the two dimensional galaxy centroiding.

\subsubsection{Gaussian Mixture Models}
Gaussian Mixture Models (GMM) offer a statistical method for grouping galaxies into substructures (similar to the method discussed in \cite{Dawson15}) by determining the best fit model for any number, N, of substructures in the system.  Increasing the number of substructures in a model almost always decreases residuals by adding more free parameters, and thus complexity.  To overcome this, we used the Bayesian Information Criterion (BIC) \citep{Liddle04}, an objective recipe for determining when the decrease in residuals is too small to justify the added complexity, when comparing models with different numbers of substructures.  We use the BIC instead of the Corrected Akaike Information Criterion (AICc) \citep{Hurvich89} when determining the most preferred model because, after testing both on a host of simulated data sets, we found that the AICc was more likely to select models with too many substructures and/or substructures placed in the wrong locations.  And while the BIC was more prone to underestimating the number of substructures present in our most realistic simulations, it did allow us to put a firm lower limit on the complexity of a system.  This is discussed in more detail in Appendix A.  We also found the use of the BIC and its ability to disfavor more complex models, despite their better fit to the data, to be especially important in our study of ZwCl 2341.1+000 when we discovered that over-splitting (dividing a distribution of galaxies into too many substructures) can also lead to an overestimation of the total dynamical mass.  This over-splitting bias is discussed further in Appendix B.

We used the previously discussed number density and luminosity maps, as well as similar maps made from the spectroscopic catalog and a convergence map made in our weak lensing analysis (this will be discussed later), to determine the models we wished to fit to the distribution of spectroscopically confirmed cluster galaxies.  For each of these models we chose the number of substructures, as well as the regions in which each substructure was confined with a uniform prior on its position.  These substructures were further constrained with uniform priors on their sizes, listed in Table \ref{tab:GMMpriors}, to ensure that we did not model non-physical clusters of galaxies.  Every model also included a large background structure to pick up stray galaxies that were not well fit to any of the substructures.  However, it was rare for more than one or two galaxies to be placed in this background.  Each of these models were run through a Markov Chain Monte Carlo (MCMC) code to determine its best fit parameters.  We then determined the most preferred model by its BIC,

\begin{equation}
BIC = \chi^2 +df(ln(n) - ln(2\pi))
\end{equation}
where $\chi^2$ is the standard measure of residuals between the model and the data, $df$ is the number of degrees of freedom (fitting parameters) in the test, and $n$ is the number of data points, or in this case, galaxies in the catalog.  The purpose of the second part of the BIC equation is to penalize more complex models for the increase in fitting parameters, thus preventing over-fitting of the data.  Each substructure added to a model increased the number of fitting parameters by eight, with three parameters for the center of the Gaussian (ra, dec, redshift), three for the variances, one for the covariance between ra and dec, and one for the Gaussian's amplitude.  We did not model any covariance between projected positions (ra and dec) and redshift because this is not a physically observed phenomenon in galaxy clusters.  For more information on this MCMC-GMM analysis, see \cite{Golovich16}.

\begin{table}
\centering
\caption{Uniform priors used in GMM to prevent the modeling of non-physically large or small clusters of galaxies}
\begin{tabular}{lll}
\hline
\hline
\multicolumn{3}{c}{Substructure Priors} \\ \hline
\multicolumn{1}{c}{Parameter} & \multicolumn{1}{c}{Min} & \multicolumn{1}{c}{Max} \\ \hline
$\sigma_{ra}$       & 0.25 Mpc  & 1.0 Mpc    \\
$\sigma_{dec}$      & 0.25 Mpc  & 1.0 Mpc    \\
$\sigma_{ra-dec}$   & 0.4 Mpc   & 1.0 Mpc    \\
$\sigma_z$       	& 300 km/s  & 1000 km/s  \\ \hline       
\end{tabular}
\label{tab:GMMpriors}
\end{table}

When comparing two models we looked at the magnitude of the difference in their BIC scores, which can be separated into four categories.  A $\Delta$BIC of 2 or less is interpreted as not significant; 2-6 as favoring the lower-BIC model but not strongly; 6-10 as favoring the lower-BIC model strongly; and above 10, as favoring the lower-BIC model very strongly \citep{Liddle04}.  Increases in the BIC can come from either an increase in the number of fitting parameters, or from an increase in the $\chi^2$.  With this, a model with more parameters can be a better fit to the data, but still obtain a higher BIC due to the penalty for its additional fitting parameters.  In such cases, we would say that the data does not support the more complex model.

\begin{figure}  
\plotone{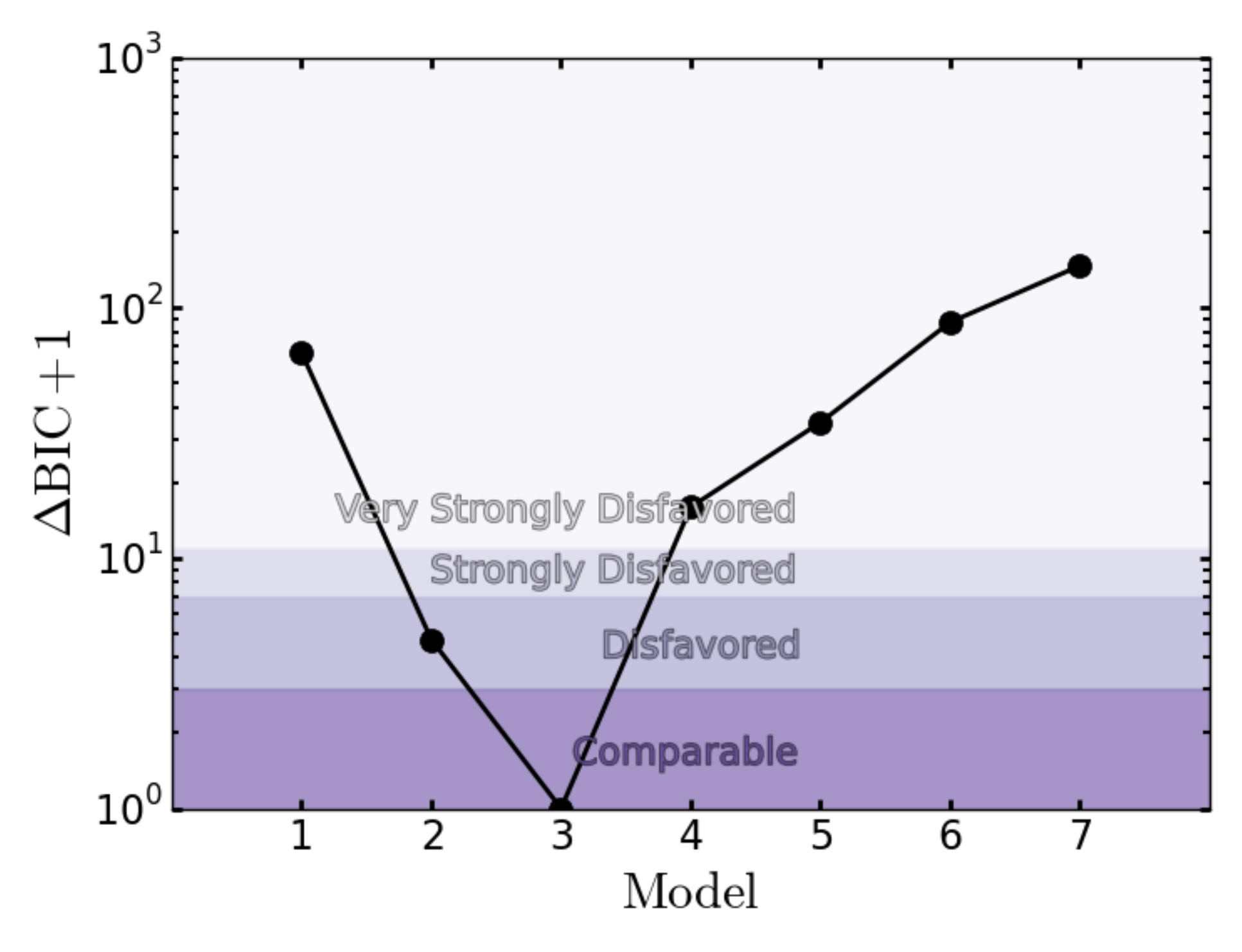}
\caption{Plot of $\Delta$BIC (+1 due to the log axis) scores for the best fit 3D GMM models relative to the most preferred model.  This most preferred model describes the system as a collection of three substructures, however, the two-substructure model is only disfavored and cannot be completely ruled out with certainty.\label{fig8}} 
\end{figure}

Figure \ref{fig8} shows that the preferred 3D GMM result is a model with three substructures.  We could not rule out the two-substructure model with certainty because it was only disfavored according to the relative BIC scores.  If we had instead used the AICc, which is less likely to underestimate the complexity of a system, the three-substructure model would still be the most preferred.  It is worth mentioning here that with 227 galaxies in our spectroscopic catalog, modeling an additional substructure increased the BIC penalty by 28.7.  Therefore, the decrease in $\chi^2$ for the three-substructure model was substantial, but barely more than the penalty for its additional fitting parameters.  Conversely, the small gains in likelihood for the four- and five-substructure models were far outweighed by the penalties for their additional fitting parameters.  Thus, we only consider the two- and three-substructure models in describing ZwCl 2341.1+0000 in the remainder of the paper.

\begin{figure}  
\plotone{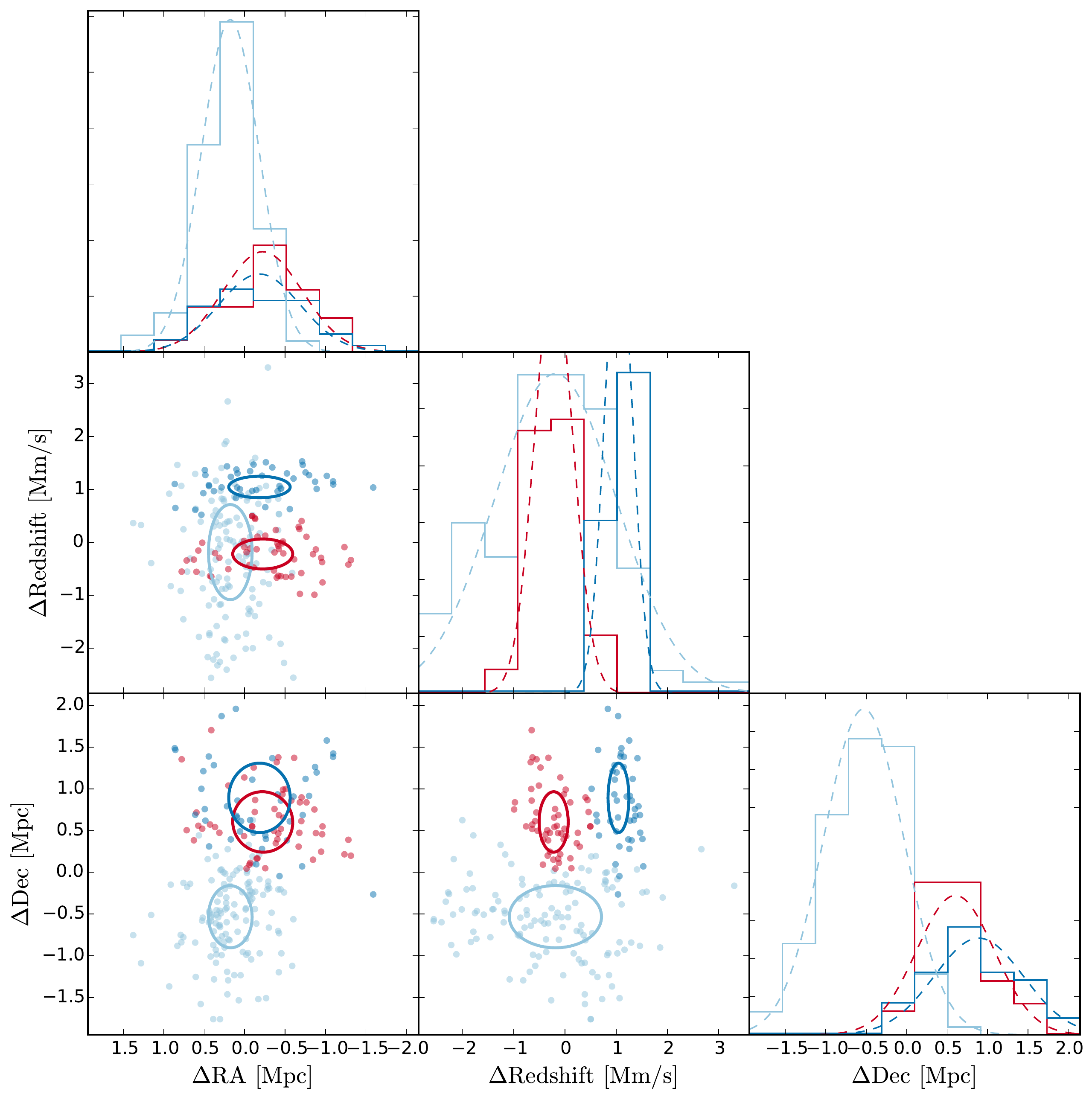}
\caption{Matrix of plots showing GMM results for the preferred three-substructure model describing ZwCl 2341.1+0000.  Boxes a, c, and f present histograms for all substructures in each of the coordinates (RA, Dec, and redshift), while boxes b, d, and e show 2D projected plots of the cluster galaxies separated into their most probable substructures.  All ellipses corresponding to clusters are drawn with 1$\sigma$ radius in each coordinate.\label{fig9}} 
\end{figure}

Figure \ref{fig9} presents the results of the three-substructure model, with panel d showing the 2D projection of the three clusters that make up the system as viewed from Earth.  This model includes one distinct cluster in the South, along with two smaller clusters in the North that are projected together in right ascension and declination space, but are separated by their line of sight velocities.  For this model we refer to these clusters as GMM-3-South (teal), GMM-3-North-A (red), and GMM-3-North-B (blue).

\begin{figure} 
\plotone{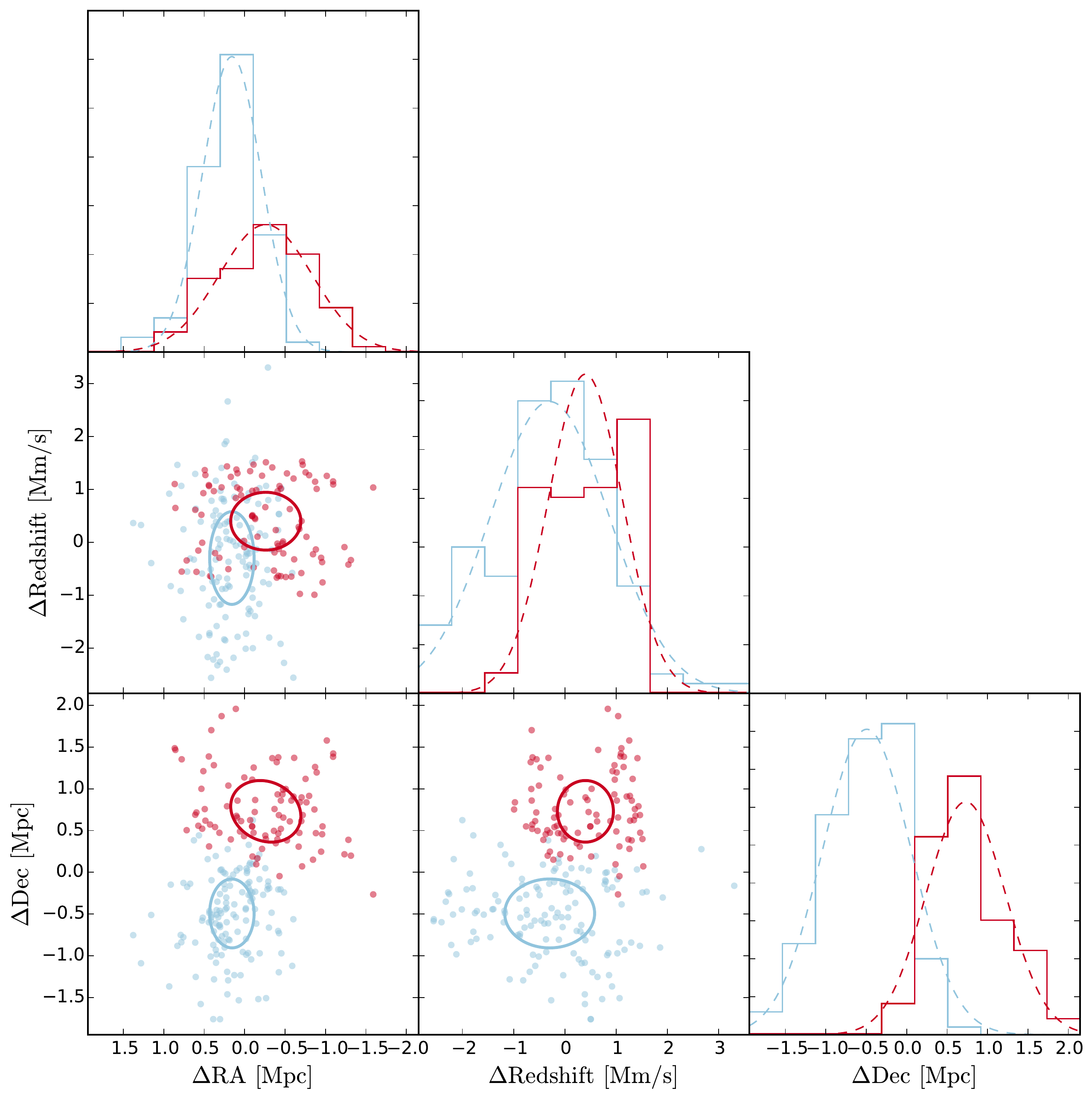}
\caption{Matrix of plots showing GMM results for the disfavored two-substructure model describing ZwCl 2341.1+0000.  Boxes a, c, and f present histograms for all substructures in each of the coordinates (RA, Dec, and redshift), while boxes b, d, and e show 2D projected plots of the cluster members separated into their most probable substructures.  All ellipses corresponding to clusters are drawn with 1$\sigma$ radius in each coordinate.\label{fig10}}  
\end{figure}

Figure \ref{fig10} presents the results of the two-substructure model.  In this model the system is split into two comparably sized clusters, which we refer to as GMM-2-South (teal) and GMM-2-North (red).

We also tested the goodness of fit for the two- and three-substructure models.  Since there is not a well established method for testing the goodness of fit for multivariate distribution models, we performed a series of one dimensional KS-tests on the projections of each substructure.  We then combined the p-values from those tests using Fisher's method \citep{Fisher48}.  This resulted in combined p-values of 0.190 and 0.233 for the two- and three-substructure models respectively, meaning that the data are consistent with both of these models.

It should be considered that the manner in which the spectroscopic data were taken may significantly bias GMM results.  There is a limit to how tightly slits could be packed on the Deimos slit masks, meaning that for each mask, dense regions tended to be under-sampled.  This could have created an incomplete catalog that was more uniform in density than the actual galaxies in the system, and could have artificially merged substructures.  It is also possible that some substructures of the system fell outside the footprint of our survey and were not sampled at all.  We did our best to mitigate these issues with our use of three slit masks, which we overlapped in the densest region of the system while also trying to cover as much area of the system as possible.  If we compare Figures \ref{fig5} and \ref{fig7}, we can see that the area enclosed in the number density contours were indeed well sampled with minimal empty regions.  However, without a spectroscopic measurement of every galaxy, we could not rule out these possible biases.  Another possibility to consider is a merger very close to pericenter with a merger axis close to the plane of the sky.  In such a scenario there would be a lot of overlap between the galaxies and no line of sight velocity difference between the two groups.  It is unlikely that our GMM analysis could accurately split the galaxies from such a scenario into their parent clusters.  For these reasons, we cannot fully rule out the possibility of there being additional substructures present in ZwCl 2341.1+0000.

\subsection{Dynamical Mass}
To infer the total dynamical mass of ZwCl 2341.1+0000 we used the virial theorem calculations of \cite{Girardi98,Girardi01} with the added use of biweights \citep{Beers90} in calculating virial quantities to decrease the importance of outlying data points.  Dynamical mass estimates for the two models previously discussed, as well as the quantities used in their calculation, are listed in Table \ref{tab:DynMass}.

\begin{table}
\centering
\begin{threeparttable}
\caption{Dynamical quantities for the clusters determined in the two- and three-substructure models as determined by the GMM analysis}
\label{tab:DynMass}
\begin{tabular}{ccccc}
\hline
\hline
\multicolumn{5}{c}{Three-Substructure Model}                                          \\ \hline
Cluster & $\bar{z}$ & $\sigma_v$ (km/s) & $R_{PV}$ (kpc) & Mass ($10^{14} M_\odot$) \\ \hline
GMM-3-South    & 0.26844  & 988   & 388     & 8.31 $\pm 0.81$ \\
GMM-3-North-A  & 0.26866  & 320   & 327     & 0.731 $\pm 0.075$ \\
GMM-3-North-B  & 0.27432  & 219   & 331     & 0.349 $\pm 0.032$ \\
Total Mass &        &       &         & 9.39 $\pm 0.81$       \\ 
\hline
\hline
\multicolumn{5}{c}{Two-Substructure Model}                                          \\ \hline
Cluster & $\bar{z}$ & $\sigma_v$ (km/s) & $R_{PV}$ (kpc) & Mass ($10^{14} M_\odot$) \\ \hline
GMM-2-South   & 0.26844  & 963    & 394     & 8.01 $\pm 0.81$  \\
GMM-2-North   & 0.27132  & 679    & 346     & 3.51 $\pm 0.34$  \\
Total Mass &	   &		&		  & 11.52 $\pm 0.87$       \\
\hline
\end{tabular}
\end{threeparttable}
\end{table}

For the three-substructure model we obtained a total dynamical mass estimate (sum of the masses of all substructures) of $9.39 \pm 0.81 \times 10^{14} M_\odot$, with less than 12$\%$ of the total mass located in the North end of the system.  The mass estimates in the North, however, are likely underestimated because GMM assigns membership to each galaxy in a way that favors minimizing the velocity dispersion where substructures are projected together (i.e. some galaxies with the highest velocity difference in GMM-3-North-A may have been assigned as galaxies near the mean velocity of GMM-3-North-B, and vice versa).  In the two-substructure model, the total mass in the North triples because all of the galaxies in the North are contained in a single, higher velocity dispersion, substructure.  But even then, the North contains less than half the mass in the South.  This increased the total dynamical mass estimate for ZwCl 2341.1+0000 to $11.52 \pm 0.87 \times 10^{14} M_\odot$, with the GMM-2-North cluster accounting for $\sim30\%$ of that.  These mass estimates are consistent with the lower mass estimates of B13 ($\sim$10$^{15} h^{-1}_{70} M_\odot$).  We expect these dynamical mass estimates to be biased high due to the disturbed nature of a merging galaxy cluster system, however they are still useful for comparing the relative masses of the substructures, assuming each substructure is similarly out of equilibrium.  In the next section we calculate a mass estimate that is independent of the dynamical state of the cluster, and also not susceptible to the over-splitting issue mentioned earlier.

\section[]{Weak Lensing Analysis}
Gravitational lensing provides the most robust form of mass estimation for merging galaxy clusters by directly probing the gravitational potential, and thus the mass, without regard for the dynamical state of the system.  In this section we present the first ever weak lensing analysis of ZwCl 2341.1+0000.  We describe the formation of a background selected weak lensing catalog, visualization of the weak lensing mass with a convergence map, and our mass estimates using the two independent techniques of aperture densitometry and model fitting.

\subsection{Weak Lensing Catalog}
The weak lensing catalog was created from our r$^\prime$ band Subaru SuprimeCam data using the magnitude selection criteria of 22.0 $\leq$ r$^{\prime}$ $\leq$ 26.1 with the added color selection of g$^\prime$ - r$^\prime$ $\geq$ -0.0833r$^\prime$ + 3.25 for galaxies brighter than r$^\prime$ = 22.5.  These selection criteria are shown on the color-magnitude plot in Figure \ref{fig6}.  We used the r$^{\prime}$ data for shape measurements because it had significantly more exposure time than the g$^\prime$ band data.  The upper magnitude limit was chosen to eliminate galaxies substantially influenced by noise fluctuations.  The other selection criteria were refined through an iterative process where we attempted to balance the purity (minimizing galaxies with z $\leq$ 0.27) and completeness (maximizing galaxies with z $\leq$ 0.27) of our catalog.  We tested the success of these cuts on a mock data set consisting of a simulated z=0.27 cluster added to the COSMOS catalog, which has precise photometric redshifts and photometry of a representative 'blank' field.  We varied the lower magnitude limit between 21.0 and 23.0, and the intercept of the color selection between 3.2 and 3.4.  We then compared the purity of the sample to the number of galaxies selected.  The purity of our mock catalog ranged between 86\% and 88\%, while the number density ranged between 36.9 to 32.5 galaxies/arcmin$^2$.  Our chosen selection criteria yielded a mock catalog with a source density of 36.2 galaxies/arcmin$^2$ and an estimated purity of $\sim$87\%.  When applied to our r$^\prime$ band Subaru SuprimeCam data, these selection criteria yielded a weak lensing catalog with a source density of 35.0 galaxies/arcmin$^2$.

In order to obtain a mass estimate for the system we had to convert reduced shear to surface mass density.  This required the distance ratio, $D_{ls}/D_s$, for source galaxies in the weak lensing catalog.  Because we did not have redshift estimations for all sources in our weak lensing catalog, we estimated a mean distance ratio ($D_{ls}/D_s = 0.409$) using our selection criteria on the same mock catalog discussed above.  We used this value throughout the weak lensing analyses.

\subsection{Convergence Map}
\begin{figure} 
\plotone{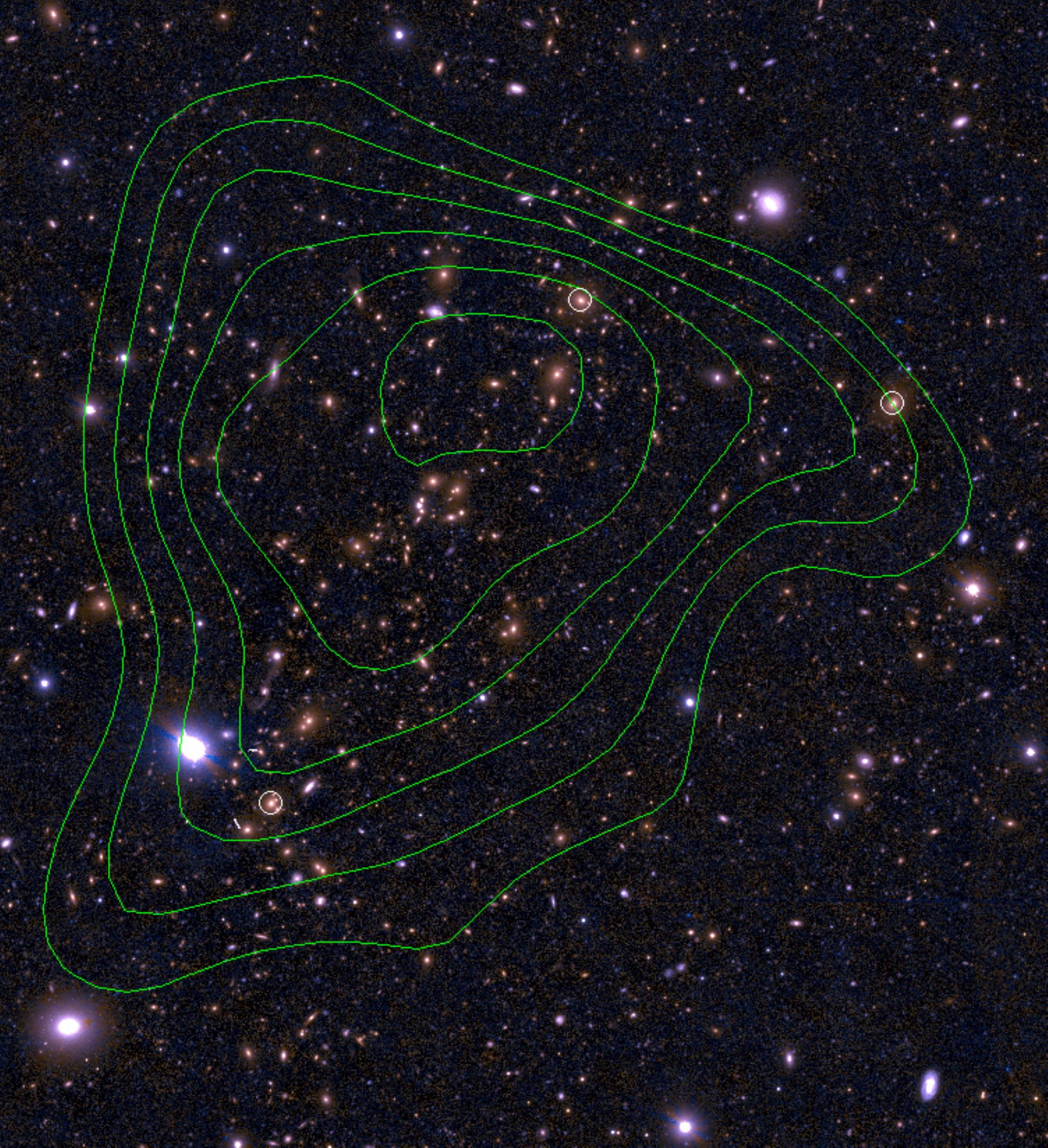}
\caption{Subaru image of ZwCl 2341.1+0000 with convergence map contours overlayed in green and the three brightest confirmed galaxies in the cluster marked with white circles.  Some of these bright galaxies behave as expected, sitting in the most dense region of the system and tracing distinct features of the convergence map, while others sit outside the contours and do not appear to be surrounded by many other, dimmer, cluster members.\label{fig11}} 
\end{figure}

In Figure \ref{fig11} we overlay the contours of a surface density (convergence) map made from the weak lensing catalog using the method of \cite{Fischer97} on the Subaru image of ZwCl 2341.1+0000.  The principal feature in the convergence map is the elongated overdensity in the central region of the cluster with the largest mass peak in the North-Center.  We also mark with white circles the three brightest confirmed red sequence galaxies in the cluster from the spectroscopic survey.  Each of these bright galaxies has an r$^\prime$ magnitude between 17.5 and 18.2.  Typically, the brightest (red sequence) cluster galaxy (BCG) for a single cluster is expected to sit in the lowest gravitational potential, and thus highest concentration of mass, in the system.  For ZwCl 2341.1+0000, it is the tenth brightest galaxy in the system that occupies this location near the center of the highest contour of the convergence map, while the second brightest galaxy is located a bit outside of this principle overdensity.  The brightest red sequence galaxy in the system is located in the North-West elongation of the convergence map, and the third brightest is located down in the South-East.  It is possible that these galaxies are the BCGs for smaller galaxy clusters/groups, whose presence helps to explain the elongations of the surface density contours.  

\subsection{Aperture Densitometry}
The simplest form of weak lensing mass estimate comes from aperture densitometry \citep{Fahlman94}.  We obtained our mass estimate using the corrected method of \cite{Clowe98, Clowe00}, which estimates the mass inside an aperture of radius $r_1$ centered at a chosen location.  This is done by calculating the zeta statistic, which measures the difference between the average convergence within radius $r_1$ and in the annulus between $r_2$ and $r_m$.  The zeta statistic is calculated by integrating over the shear components tangential to the chosen aperture center for all source galaxies, as shown by Equation 2.  

\begin{align}
\zeta(r_1,r_2,r_m) 
&= 2 \int_{r_1}^{r_2} \frac{1}{r}<\gamma_t>dr + \frac{2}{1-r_2^2/r_m^2}\int_{r_2}^{r_m} \frac{1}{r}<\gamma_t>dr \\
&= <\kappa (r_1)> - <\kappa (r_2,r_m)>
\end{align}

If the chosen $r_1$ is large enough to contain all of the cluster mass, then the mean convergence within the outer annulus should go to zero.  This equates the zeta statistic to the mean convergence inside the aperture, which can be used to calculate an estimate for the total mass contained within said aperture, as shown in equation 4.

\begin{equation}
M(<r_1) = 2\pi r_1^2\Sigma_{cr} <\kappa (r_1)> = 2\pi r_1^2\Sigma_{cr} \zeta(r_1,r_2,r_m)
\end{equation}

\begin{figure}  
\plotone{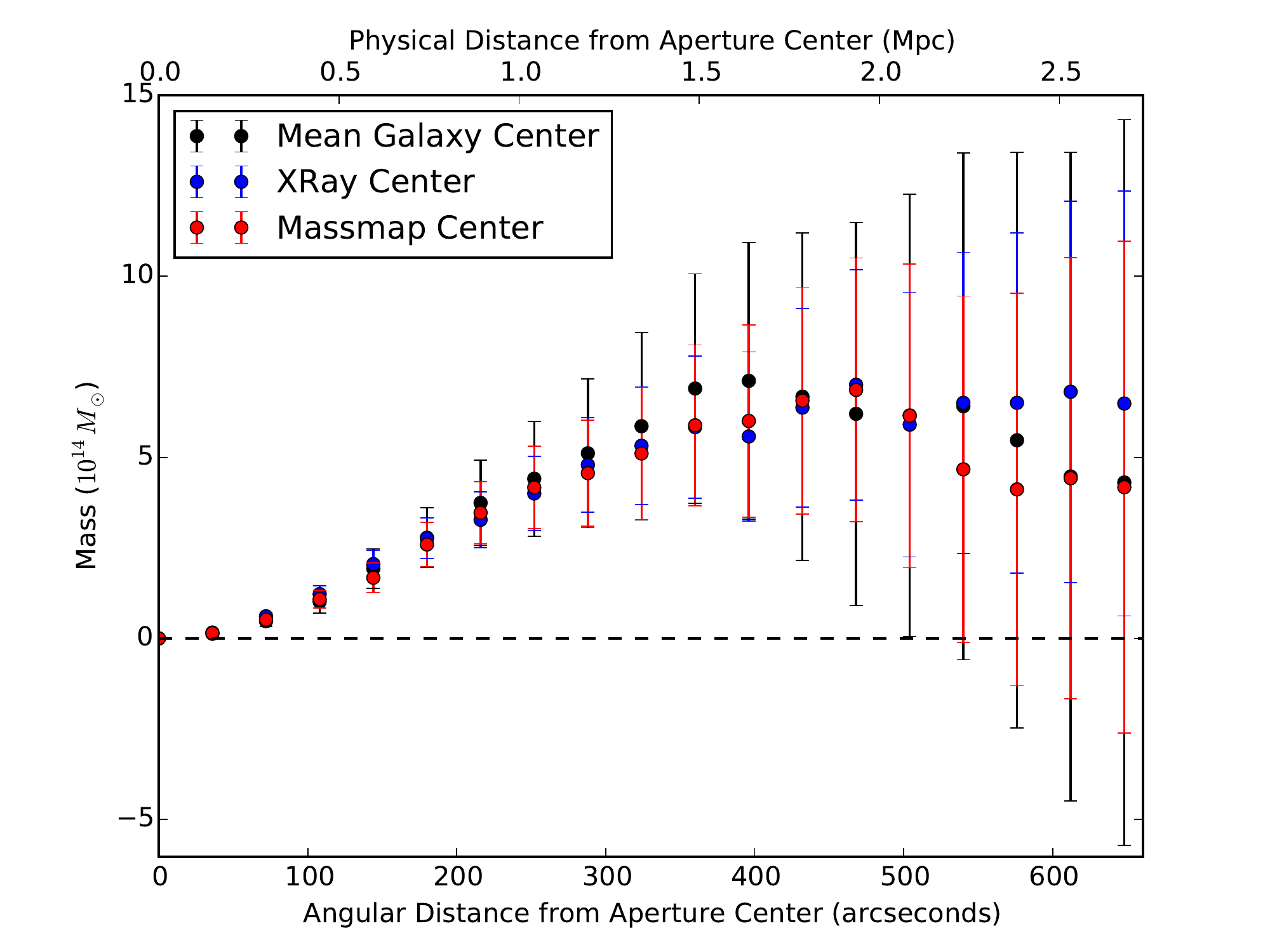}
\caption{Aperture mass estimates for ZwCl 2341.1+0000 in successively larger apertures for the three selected aperture centers.  Blue dots represent an aperture centered on the peak X-ray emission, black dots represent an aperture centered on the mean location for spectroscopic confirmed cluster members, and red dots are for the aperture centered upon the main peak in the convergence map.\label{fig12}} 
\end{figure}

However, this mass estimate assumes axisymmetry, which ZwCl 2341.1+0000 violates.  Testing on simulations showed that aperture densitometry tends to over-estimate the weak lensing mass of asymmetric systems, but could still give fairly accurate masses for mass distributions with a single distinct mass peak.   Therefore, we can still use this as an approximation on the total weak lensing mass of ZwCl 2341.1+0000.  Having an asymmetric system also introduces a complication in choosing a location to center the aperture upon.  We decided to use three separate centerings for our apertures, the first of which is the main X-ray emission peak, which should indicate the bulk of the ICM.  Our other two centering locations are the mean galaxy position from the spectroscopic catalog, and the center of the largest overdensity in the convergence map.  For each of the centerings, most confirmed cluster galaxies are contained within the aperture when the inner radius reaches 400 arcseconds ($\sim$1.6 Mpc).  Such an aperture size is around what we would expect to contain the majority of the cluster mass as well, since, using the \cite{Duffy08} relations, r$_{200}$ for a $10^{14} M_\odot$ ($10^{15} M_\odot$) galaxy cluster is $\sim$210 arcseconds ($\sim$450 arcseconds) at the cluster redshift.  This can also be seen in Figure \ref{fig12}, which presents the mass estimates for each aperture size used with all three centerings.  Apertures larger than 400 arcseconds yield roughly the same mass estimate, but with larger uncertainties.  For all three of the centerings, every confirmed cluster galaxy is well contained inside of an aperture with a radius of 600 arcseconds ($\sim$ 2.5 Mpc).  We chose $r_2$ = 750 arcseconds and $r_m$ = 825 arcseconds so that the outer annulus included as little (if any) cluster mass as possible, but also did not lie too close to the edge of our survey field.  Results for all three centerings were consistent with a mass of $\sim 6 \times 10^{14} M_\odot$ with uncertainties of $\sim 4 \times 10^{14} M_\odot$.

Something to note here is that we observe the reduced shear (g), and not the shear ($\gamma$) directly.  These two quantities are related by:

\begin{align}
g = \frac{\gamma}{1-\kappa}
\end{align}
If $\kappa$ is sufficiently small, then the shear and reduced shear can be roughly equated to each other.  For a $10^{15} M_{\odot}$ galaxy cluster at a redshift of 0.27, $\kappa$ becomes sufficiently small ($<$ 0.01) at a distance of 380 arcseconds.  Because we are looking at a cluster with a mass less than $10^{15} M_{\odot}$ and are only considering the estimates at more than 400 arcseconds from the cluster center as viable mass estimates of the system, we can safely assume we are working in the weak lensing regime where g $\approx \gamma$.

\subsection{Multiple Halo Model Fitting}
For a more precise weak lensing mass estimate of ZwCl 2341.1+0000 and a chance to better describe the mass distribution in the system, we used a series of model fitting analyses, which we refer to as Multiple Halo Model Fitting (MHMF).  For this method we modeled a number of NFW profile halos using the \cite{Duffy08} mass-concentration relationship at the cluster redshift.  We then calculated reduced shear components as a linear combination of the shear produced by each modeled halo for every galaxy in our weak lensing catalog.  These reduced shear components were compared to the actual measured shapes of the source galaxies and a residual difference was calculated.  This process was repeated in a least squares fit where the mass and position (right ascension and declination) of each modeled halo were free to change until a best fit was obtained.  

MHMF was performed with models containing between zero and six halos.  The zero mass model was our null test, where we assumed that there was no excess matter present at the cluster redshift, and that the shapes of the source galaxies in our weak lensing catalog were randomly oriented.  All of the other models tested were motivated by the work already discussed, such as the peaks and elongations present in the convergence, projected number density, and luminosity maps, and the centers of substructures determined by the GMM clustering.  The most preferred models were again determined by BIC score as described in \S 4.1.1.  

\begin{figure}
\plotone{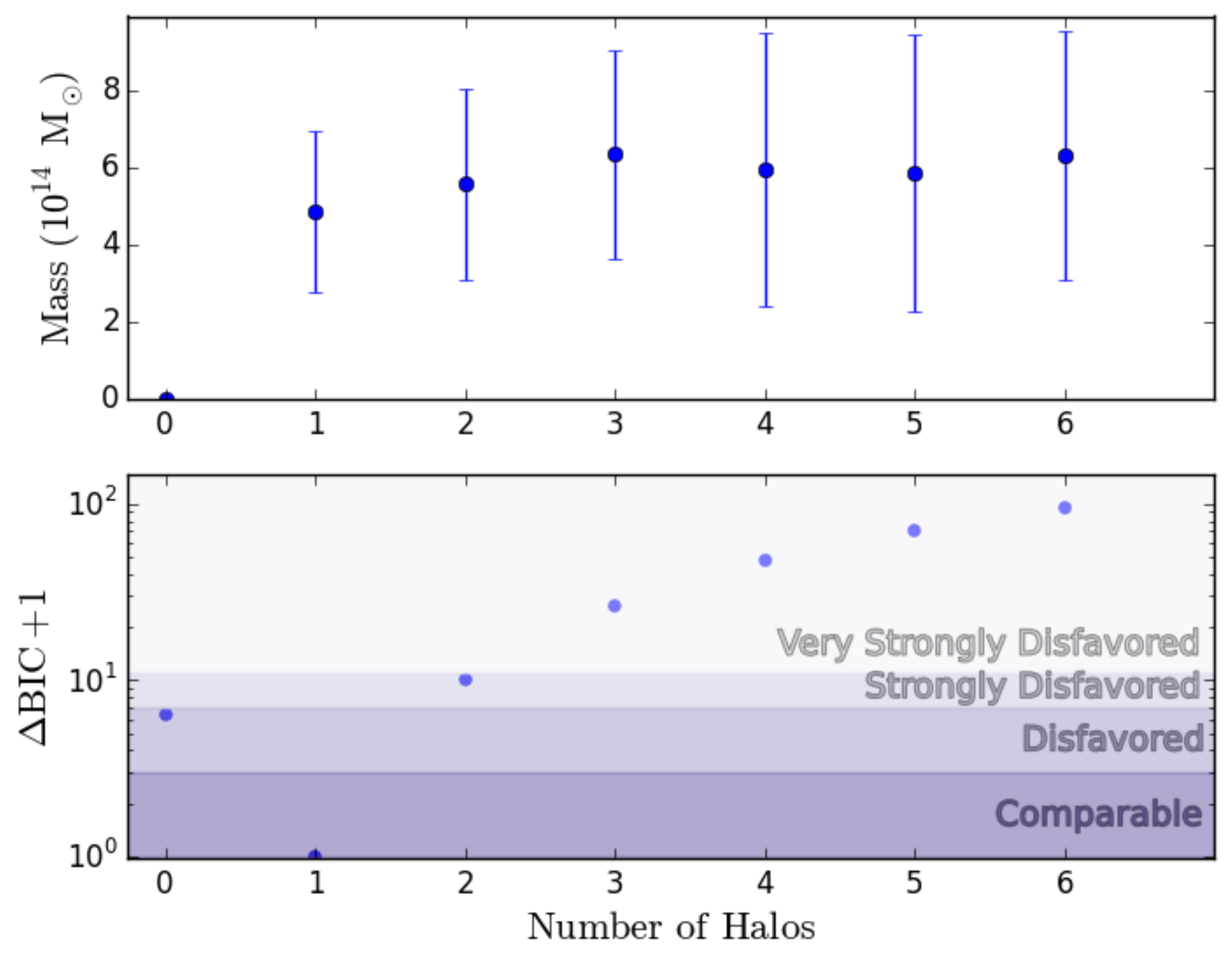}
\caption{Top plot presents the mass estimates for the best performing models for each number of halos modeled with MHMF.  The bottom plot shows the $\Delta$BIC (+1 due to log axis) comparison for each of these weak lensing models tested, with the one-halo model clearly preferred over the others.\label{fig13}} 
\end{figure}

Figure \ref{fig13} presents the total mass estimates (sum of masses for all halos in the model) along with the $\Delta$BIC scores (relative to the best fit model) of the best performing models for each number of halos tested.  In Table \ref{tab:MHMFMass} we show the results for the best performing one-halo and two-halo models.  The uncertainties on these results are from the model fitting alone, and are likely underestimated.  We found that by changing the weak lensing catalog (i.e. using a different selection criteria for a different balance of purity and completeness) the positions of the halos in the favored model could change, in some cases by more than 3$\sigma$ from the results shown here.  The mass distribution and total mass estimate for the system, however, were not significantly altered when using the different catalogs.

\begin{table*}
\centering
\caption{Results for the best performing one- and two-halo models from MHMF, listing the positions (right ascension and declination) and Mass ($M_{200}$) for each halo, as well as the total cluster mass for each model}
\label{tab:MHMFMass}
\begin{tabular}{llll}
\hline
\hline
\multicolumn{4}{c}{One-halo Model}                                          \\ \hline
Halo   & $\alpha$ (h:m:s) & $\delta$ ($^\circ$ ' ") & $M_{200}$ ($10^{14} M_\odot$) \\ \hline
WL-1 & 23:43:42.53 $\pm$ 0:0:0.98 & 0:19:18.37 $\pm$ 0:0:19.37 & 4.86 $\pm$ 2.09 \\ \hline
\hline
\multicolumn{4}{c}{Two-halo Model}                                          \\ \hline
Halo   & $\alpha$ (h m s) & $\delta$ (d m s) & $M_{200}$ ($10^{14} M_\odot$) \\ \hline 
WL-2-South & 23:43:46.89 $\pm$ 0:0:0.46 & 0:16:34.56 $\pm$ 0:0:8.53 & 2.74 $\pm$ 1.73 \\
WL-2-North & 23:43:39.58 $\pm$ 0:0:3.13 & 0:20:22.03 $\pm$ 0:0:34.02 & 2.83 $\pm$ 1.77 \\
Total Mass & & & 5.57 $\pm$ 2.47 \\
\hline
\end{tabular}
\end{table*}

\begin{figure}
\plotone{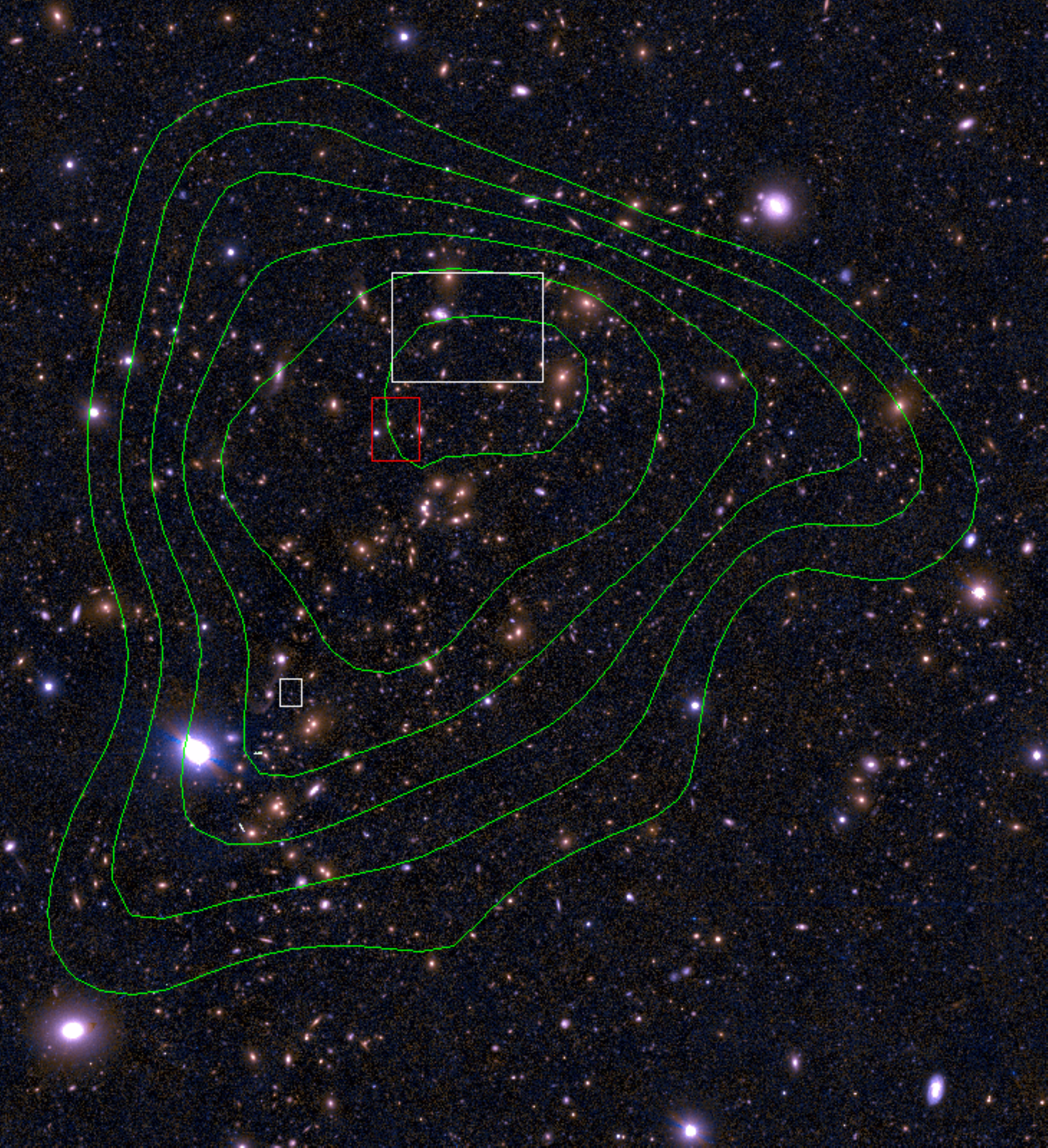}
\caption{Subaru color image of ZwCl 2341.1+0000 with surface mass density contours (green) and the positions of the MHMF best fit halos marked.  The position of the one-halo model is marked with a red box, while halos for the two-halo model are marked in white.  Box sizes indicate the uncertainty in RA and Dec positions for each halo.\label{fig14}} 
\end{figure}

The most preferred model was a single halo centered at $RA = 23^h 43^m 42.53s \pm 0^h 0^m 0.98^s$, $Dec = 0^\circ 19' 18.37" \pm 0^\circ 0' 19.37"$, with a weak lensing mass estimate of $4.86 \pm 2.09 \times 10^{14} M_\odot$.  Here we should note that with our weak lensing catalog, each additional halo added to a model increased the BIC score by 25.6.  This means that even though the one-halo model is the most preferred via BIC score, the two-halo model was a significantly better fit to the data.  We consider this two-halo model to be a better description of the system despite its higher BIC score because of the evidence from spectroscopy supporting more than one substructure in ZwCl 2341.1+0000, the presence of two radio relics indicating a merger between at least two galaxy clusters, and because of the low signal to noise ratio (S/N $<$ 3) in our weak lensing data, which could have made it impossible to favor any of the more complex models.  This two-halo model predicted a relatively equal split of the system's mass between the North and South, but the large uncertainties do not preclude substantially different masses from these central values.  The positions of the halos from both models (with uncertainties) are shown in Figure \ref{fig14}.

We should also note that when testing MHMF with simulations matching the data quality in our weak lensing catalog, we found it difficult to distinguish two halos when they were placed too close together.  In such simulations, the preferred model was typically a single halo that accurately predicted their combined mass and was placed between the two simulated halos.  Increasing the mass ratio of the two simulated halos brought the position of the modeled halo closer to the simulated halo with more mass.  Thus, our two-halo model could have underestimated the number of halos present in the system if some were projected very close together, as the GMM three-substructure model predicted in the North.

Despite the differences in the models, both in number of halos and where those halos were located, all weak lensing total mass estimates for ZwCl 2341.1+0000 were within one sigma uncertainty of each other.  This robustness of the weak lensing mass estimate was also tested and confirmed on a host of simulated data sets with known mass and varying signal to noise ratio, showing that unlike its dynamical counterpart, MHMF is not very susceptible to changes in the total mass estimate due to how the cluster is divided.  

\section{Discussion and Conclusions}
The smoothed number density map of ZwCl 2341.1+0000 presented by B13 appeared to contain a large number of overdense regions of galaxies, which could indicate a large number of galaxy clusters in the system.  This information, coupled with the presence of an elongated X-ray emitting gas cloud bracketed by two radio relics, and B13's own analyses, led them to the conclusion that ZwCl 2341.1+0000 was a complex and massive system consisting of multiple merging galaxy clusters.  Our own analyses agree with B13's on the potential complexity of ZwCl 2341.1+0000.  However, we found the system to be considerably less massive than previously estimated.

From our GMM analysis we found that a model with three substructures most economically described the distribution of spectroscopically confirmed cluster galaxies, but a simpler two-substructure model could not be ruled out with confidence.  Both of these models included a cluster in the South, which was coincident with the main luminosity peak of the system (Figure \ref{fig7}, Right), as well as the Southern elongation of the convergence map (Figure \ref{fig11}), and the WL-2-South halo from the two-halo weak lensing model (Figure \ref{fig14}).  The three-substructure model also included two much less massive clusters in the North, which had small projected separations, yet were radially separated by a $\sim$1300 km/s mean velocity difference.  We were able to distinguish these two Northern clusters from a single distribution due to our larger spectroscopic sample size.  In the disfavored two-substructure GMM model, the galaxies in the North were not divided in velocity space, but were instead kept as a single, more massive cluster.  The positions of these Northern clusters were consistent with the location of the WL-2-North halo from the MHMF two-halo model, as well as the main peak in the convergence map and the Northern peak in the luminosity map.

\begin{figure} 
\epsscale{0.7}
\plotone{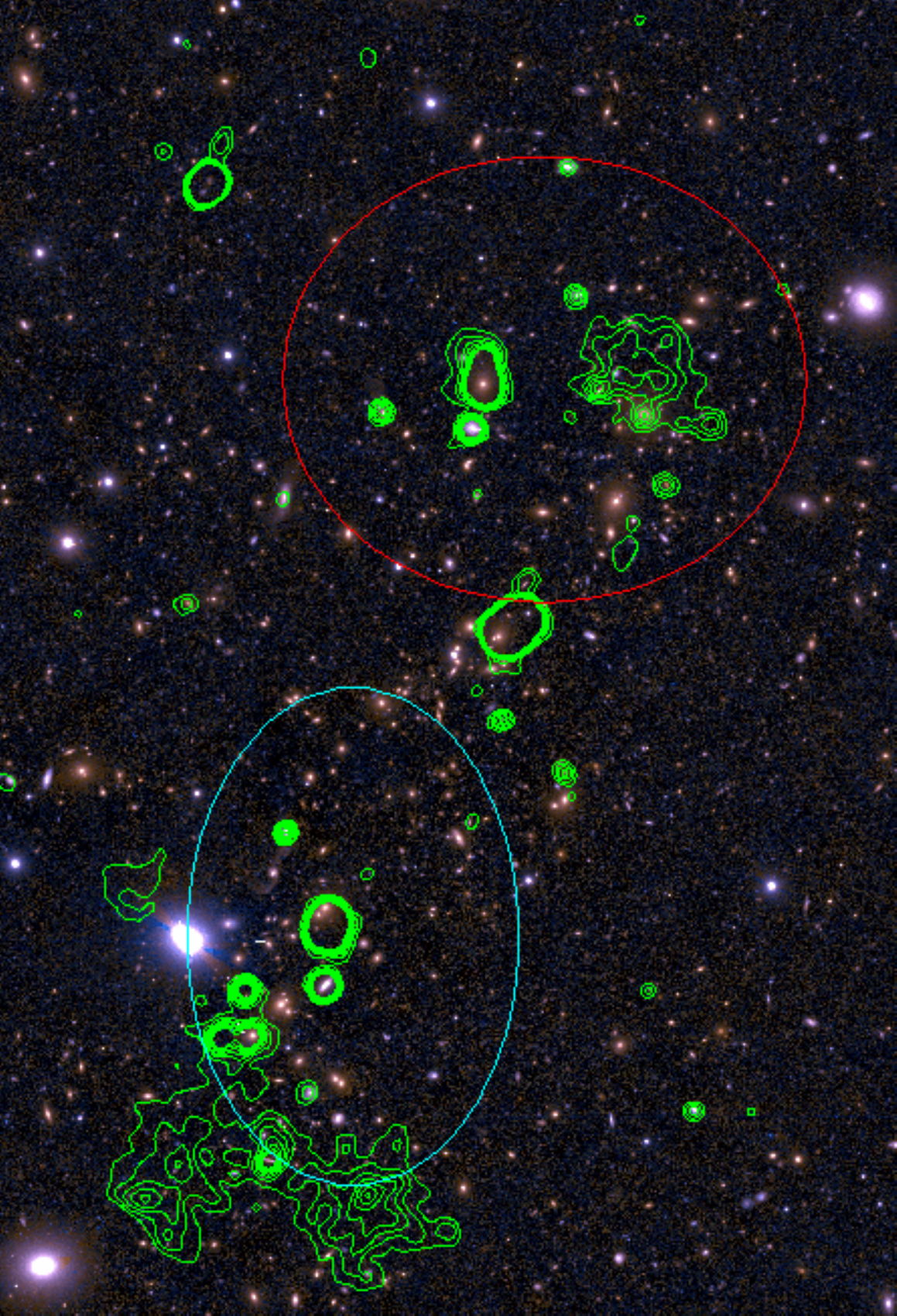}
\caption{Subaru color image of ZwCl 2341.1+0000 with GMRT radio contours (green) GMM substructure results for the GMM-2-South (teal) and GMM-2-North (red) clusters overlayed.  The diffuse radio emissions in the South-East indicate the South radio relic, which coincides as we would expect with the GMM-3-South cluster.  The diffuse radio emissions in the North-West indicate the North radio relic, which is not appropriately located with respect to the GMM-2-North cluster if it was generated by a merger event involving the GMM-2-South cluster.\label{fig15}}  
\end{figure}

In Figure \ref{fig15} we present a Subaru image of ZwCl 2341.1+0000 marked with ellipses for the clusters from the disfavored two-substructure GMM model along with the GMRT radio contours.  Here we can see that the South relic is leading the GMM-2-South cluster as they travel away from the center of the system, as we would expect in a typical merger between GMM-2-South and GMM-2-North.  However, in this model the North relic lags behind a large number of the Northern galaxies and is situated closer to the center of the GMM-2-North cluster, rather than its leading edge.  While it is not impossible for a shock to lag behind some of the cluster galaxies, it is also not common.  Furthermore, the large separation between the two clusters typically indicates a long time since pericenter, and that the merger is occurring near to the plane of the sky.  Both factors suggest that the shock should not be lagging behind the cluster galaxies.

\begin{figure}
\epsscale{0.7}
\plotone{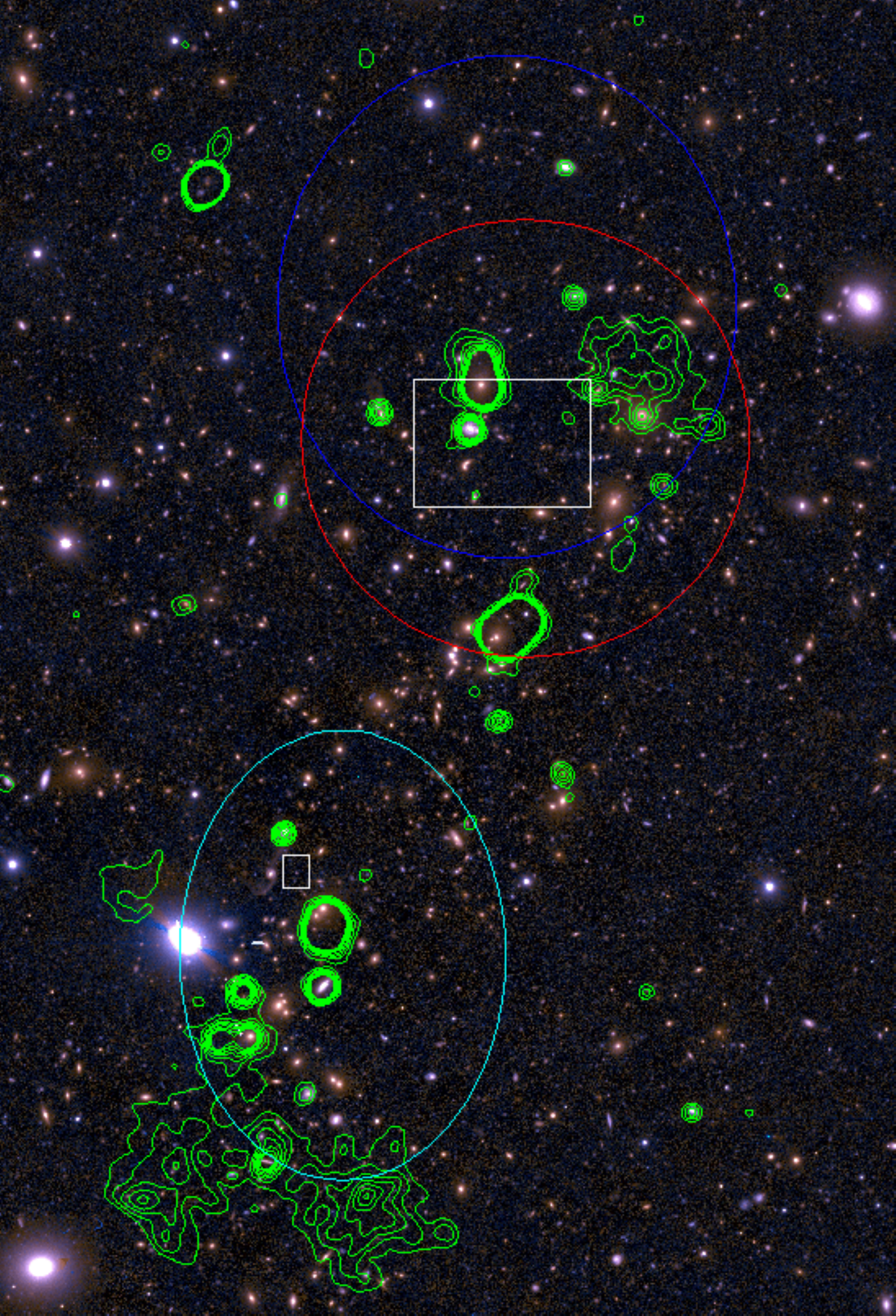}
\caption{Subaru color image of ZwCl 2341.1+0000 with GMRT radio contours (green), GMM substructure results for the GMM-3-South (teal), GMM-3-North-A (red) and GMM-3-North-B clusters, and the MHMF two-halo weak lensing mass locations (white) overlayed.  The diffuse radio emissions in the South-East indicate the South radio relic, which coincides as we would expect with the 3-South cluster and WL-2-South positions.  The diffuse radio emissions in the North-West indicate the North radio relic, which is consistent with a merger between GMM-3-South and GMM-3-North-A.\label{fig16}} 
\end{figure}

Figure \ref{fig16} similarly shows the GMRT radio contours along with ellipses for the three clusters from the preferred GMM result and boxes for the two-halo MHMF result overlayed on a Subaru color image of ZwCl 2341.1+0000.  If we first consider the GMM results, we again see that the South relic is leading the GMM-3-South cluster as they travel away from the center of the system, as we would expect in a typical merger between GMM-3-South and one of the GMM-3-North clusters.  From its position relative to the South relic, the North relic appears to be leading the densest collection of galaxies associated with the GMM-3-North-A cluster as they also travel away from the center of the system.  Because the North relic is leading the bulk of GMM-3-North-A galaxies, we do not consider it to be an issue that the relic lies slightly inside the GMM-3-North-A 1-sigma radius.  This scenario explains the shock location more cleanly than the two-substructure model does.  It is also possible that the North relic was generated by a merger between the two clusters in the North.  However, the position of the North radio relic is more consistent with the merger axis defined by GMM-3-South and GMM-3-North-A than the one between GMM-3-North-A and GMM-3-North-B, as relics are usually not found between two merging galaxy clusters.  

Looking at the MHMF results, we can see that the positions of the North and South radio relics are well described by a merger between WL-2-South and WL-2-North.  The position of the northern lensing mass also best matches the position of GMM-3-North-A, which fits with our scenario where GMM-3-South and GMM-3-North-A produced the relics.  With our low-resolution weak lensing, we expect the the mass we detected in the North to be a combination of GMM-3-North-A and GMM-3-North-B.  It is possible that GMM-3-North-A contains more mass (it does have the brighter BCG and the higher dynamical mass estimate), than GMM-3-North-B, and thus could have more weight in determining the combined weak lensing position.  Because of this low resolution in our weak lensing data, we do not consider there to be any tension between the GMM and MHMF results on the number of clusters present in the North.  

There does appear to be some tension between the GMM three-substructure model and the weak lensing results when we compare their distributions of the mass in the system.  According to the latter, the mass should be evenly divided between the North and the South, but our dynamical mass estimate for the three-substructure model places significantly more mass in the South than the North.  This tension can be alleviated somewhat with the two-substructure model, which increases the mass in the North by not splitting the Northern galaxies.  However, for the reasons already discussed, we do not consider this to be correct.  We therefore compare the weak lensing mass of WL-2-North with the sum of the dynamical masses of GMM-3-North-A and GMM-3-North-B.  This combined dynamical mass is, remarkably, less than than its weak lensing counterpart, which is not something we expect in a merging system.  However, this disagreement is not much more than one sigma, as unity only requires a one sigma decrease in the weak lensing mass without any upward shift of the dynamical mass.  Furthermore, as we mentioned earlier, the two dynamical masses in the North are likely underestimated due to how GMM clusters the galaxies.  In contrast, the dynamical mass estimate for GMM-3-South is $\sim$3.0 ($\pm$ 1.9) times  the weak lensing mass of WL-2-South.  Since it has been shown that a merger can bias the dynamical mass estimate high by as much as a factor of four \citep{Pinkney96}, we do not consider this to be an issue at all.  We also note here that the two clusters participating in a merger do not have to experience equal disruptions to their galaxy velocities (dynamical mass).  In which case, our results could indicate that GMM-3-South has experienced more dynamical disruption due to the merger than GMM-3-North-A.

Considering the merger history of ZwCl 2341.1+0000, we can see that the primary merger event has occurred along the North-West to South-East axis, and was probably responsible for generating both of the observed radio relics.  Such a merger is best described by the GMM-3-North-A and GMM-3-South clusters from the GMM analysis, which correspond well with the WL-2-North and WL-2-South weak lensing halos found with MHMF.  Following the methods described in \cite{Dawson13}, we used the mass estimates from the two-halo weak lensing model, along with the positions and velocity distributions of the clusters in the GMM three-substructure model and the the integrated polarization fractions of the radio relics, to constrain some of the merger properties in an attempt to better reconstruct the primary merger of ZwCl 2341.1+0000.  The weak lensing mass in the North was divided proportionately between the GMM-3-North-A and GMM-3-North-B clusters based on the ratio of their dynamical mass estimates.  However, the exact division of this Northern weak lensing mass turned out to be unimportant due to its large uncertainty, which is taken into account in the reconstruction.

From the merger reconstruction, we were most interested in the merger axis angle, which gives the orientation with respect to the plane of the sky along which the merger is taking place, the collision speed, which is the relative speed of the two substructures at pericenter, and the time since collision, which is how long it has been since the clusters crossed at pericenter.  There is a degeneracy in the time since collision calculation, because it is possible that two clusters could have the same observed relative positions and motions and either still be moving away from each other, or be coming back towards each other for a second core crossing after having reached maximum separation.  We calculate both of these values in reconstructing a merger and use other information, such as a the positions of the radio relics relative to the clusters, to help break the degeneracy.  It is important to note here that all of our calculations only use information from two clusters at a time, as our analysis was designed to reconstruct a bimodal merger and was not equipped to handle a more complex system.  More precise values could be calculated by accounting for all of the clusters simultaneously, however this is outside the scope of the current paper.  As such, we take the following values discussed to be rough estimates.

Using the values discussed earlier, the \cite{Dawson13} algorithm finds the merger axis angle between GMM-3-South (WL-2-South) and GMM-3-North-A (WL-2-North) to be $\sim$10$^{+34}_{-6}$ degrees, and the collision speed at pericenter to be $\sim$1900$^{+300}_{-200}$ km/s.  Considering the small projected separations between the relics and their associated clusters, we believe this to be a scenario where the two clusters are still moving apart.  In which case, we estimate the time since collision to be $\sim$1.1$^{+0.7}_{-0.3}$ Gyr.  If the clusters are instead moving back together, our best estimate for the time since collision would be $\sim$1.5$^{+1.4}_{-0.3}$ Gyr. The large uncertainties here, especially for the merger axis angle, are primarily due to the large uncertainties on the weak lensing masses.  With better weak lensing measurements, we could reduce those uncertainties considerably.

We were unable to place any meaningful constraints on the parameters describing the mergers between any other pairs of galaxy clusters in the system due to the large uncertainties in their masses and positions, as well as the lack of any radio relics associated with their merging.  It is even possible that some pairs of clusters have not yet experienced their first core crossing, and are still going through their first in-fall.  This could account for the lack of other radio relics.  However, that is not definitive, as radio relics are not always observed when two galaxy clusters merge.  Despite being unable to constrain many of the merger parameters, our dynamical modeling did show that ZwCl 2341.1+0000 fits comfortably into the general understanding of merging clusters with radio features.

In conclusion, we found that ZwCl 2341.1+0000 was likely a complex merger comprised of at least three galaxy clusters, and was less massive than previously estimated.  With our current data, a three-substructure model was preferred in describing the system.  The two-substructure model was viable as an explanation of the spectroscopic data alone and created less tension with the weak lensing mass distribution, but created more tension with the radio data by placing the northern relic in the middle of the northern substructure.  With more redshift measurements of galaxies in this region we could increase our certainty against the simpler model, and potentially see a split in the galaxies in the Southern half of the system.  The division of the Northern galaxies into two separate clusters did decrease the total dynamical mass estimate for the system.  However, the weak lensing mass estimate obtained through model fitting was even lower still ($5.57 \pm 2.47 \times 10^{14} M_{\odot}$), and should be a more accurate measurement of the total mass in ZwCl 2341.1+0000.  While our best fit weak lensing model contained only one NFW profile mass halo, we considered the two-halo model to be a much better description of the system given all of the other information available.  The low signal to noise ratio in our weak lensing data left us with very large uncertainties, and was likely responsible for our inability to rule out the simpler models.  More exposure time in the optical r' band could increase the signal to noise ratio of our weak lensing data and allow us to better model the mass distribution in ZwCl 2341.1+0000.  Similarly, a more detailed radio study of the system would be required to resolve the somewhat unusual properties (i.e. potentially low polarization fraction) of the radio relics present.

\acknowledgments
This research has made use of the galaxy catalogue of the Sloan Digital Sky Survey (SDSS). The SDSS web site is http://www.sdss.org/, where the list of the funding organizations and collaborating institutions can be found.  This research has also made use of the galaxy catalogue of the Cosmological Evolution Survey (COSMOS).  The COSMOS web site is cosmos.astro.caltech.edu, where the list of the funding organizations and collaborating institutions can be found.  DW and BB acknowledge support from NSF grant 1518246.  Part of this was work performed under the auspices of the U.S. DOE by LLNL under Contract DE-AC52-07NA27344.  M.J.J. acknowledges support from the National Research Foundation of Korea to the Center for Galaxy Evolution (CGER).

{\it Facilities:} \facility{Subaru (Suprime-Cam)}, \facility{Keck:II (DEIMOS)}

\appendix

\section{Comparing AICc and BIC Success in Model Selection Using GMM}
There are many methods available for model selection.  However, the Akaike and Bayesian Information Criterion (AIC and BIC respectively) are the most popular when selecting between models with different numbers of free parameters.  While both of these are prior independent, the BIC also has the advantageous property of consistency.  For a criterion, consistency means that is provably immune to false positives in the asymptotic limit, and will always select the ``true" model if it is among those being tested.  If the ``true" model is not among those being tested, the BIC will instead choose the most parsimonious model that is closest to the ``true" model \citep{Claeskens08}.  The AIC, on the other hand, is asymptotically efficient, meaning that it will always select the model which minimizes the mean squared error of prediction.  In the case of the AIC, the error of prediction is the expected Kullback discrepancy \citep{Cavanaugh11}.  In our tests, we chose to use the Corrected Akaike Information Criterion (AICc), which is the AIC with a correction for finite sample size that effectively increases its penalty for added free parameters.  The free parameter penalty of the BIC, however, is still greater than that of the AICc.

In testing the efficacy of these two criteria for selecting the best model with our GMM clustering analysis, we produced mock spectroscopic catalogs that matched the spatial volume and data quality of our true spectroscopic catalog for ZwCl 2341.1+0000.  The was done by selecting the number of substructures to simulate, then randomly generating the central positions (Right Ascension, Declination, and mean radial velocity) for each substructure within the confined volume.  A mass was then randomly selected for each substructure ranging between 0.5 to 20.0 $\times 10^{14} M_{\odot}$, and a virial projected scale radius and velocity dispersion were calculated using the methods of \cite{Girardi01}.  Galaxies for each substructure were then generated in one of four ways, giving us four different categories of simulations.

The first, and simplest, category we refer to as Gaussian simulations.  Here we divided the total number of galaxies to be generated unevenly between the number of substructures in the simulation.  We then simulated each substructure using a two-dimensional Gaussian for the projected positions, and a one-dimensional Gaussian for the velocities, of its galaxies.  The second category, which we refer to as Gaussian Masked, generated between 500 and 1000 galaxies for each substructure using the same Gaussian distributions mentioned above.  A grid was then applied to the projected RA-Dec space of the simulated catalog, with the length and width of each grid cell equal to the average slit length on a DEIMOS mask.  We then randomly selected 225 galaxies from the catalog with the added restriction that a maximum of one galaxy could be selected from each cell.  This was done to mimic the more regular distribution galaxies created by the restrictions of using a DEIMOS slit mask to take spectroscopic measurements without going through the time consuming process of creating actual slit masks for the simulated data sets.  For both of these categories we tested ten simulations for each number of substructures simulated, which ranged from one to four.

The last two categories of simulations are referred to as Student-t, and Student-t Masked.  These are very similar to the first two, with the exception that the projected positions of the galaxies were generated using a two-dimensional Student-t distribution rather than a Gaussian.  We again tested the same number of simulations for the Student-t category as the two above.  Because the true distribution of galaxies in a cluster is also non-Gaussian, we believe the Student-t Masked simulations are the most effective test of our GMM analysis, and of the better criterion for selecting the most preferred model.  As such, we generated and tested substantially more simulations in this category, with 110 simulations tested for each number of substructures used.

Each simulated catalog was treated in the same manner as the true spectroscopic catalog of ZwCl 2341.1+0000 discussed in this paper (i.e. visualizations of the galaxy distribution were made and inspected, overdensities located, and priors on the positions for substructures to be analyzed were chosen by hand).  During this process the true positions of the simulated substructures were kept blind, and only looked at after the analysis was completed.  All simulations were tested for one to six substructures using our GMM analysis discussed in \S 4.1.1.  A summary of the results for these tests are listed in Tables \ref{tab:GMMResults1} and \ref{tab:GMMResults2}.

\begin{table}
\centering
\caption{Summary of results for the Gaussian and Gaussian Masked simulations used to test efficacy of BIC and AICc in selecting the most appropriate model using our GMM analysis.  For each category, the number of simulated substructures in the mock catalog is listed down the left side, and the number of substructures in the preferred model is listed along the top.  In each cell of the table we list the number of simulations that yielded a preferred model with that number of substructures using BIC or AICc.}
\label{tab:GMMResults1}
\begin{tabular}{c|cccc|cccc}

\hline
\hline
\multicolumn{9}{c}{Gaussian} \\ \hline
 & \multicolumn{4}{c|}{BIC} & \multicolumn{4}{c}{AICc}\\ \hline
 & 1 & 2 & 3 & 4 & 1 & 2 & 3 & 4 \\ \hline
1 substructure  & 7 & 3 & 0 & 0 & 5 & 2 & 3 & 0 \\
2 substructures & 0 & 9 & 1 & 0 & 0 & 9 & 1 & 0 \\
3 substructures & 0 & 5 & 5 & 0 & 0 & 3 & 6 & 1 \\
4 substructures & 0 & 0 & 5 & 5 & 0 & 0 & 4 & 6 \\
\hline
\hline
\multicolumn{9}{c}{Gaussian Masked} \\ \hline
 & \multicolumn{4}{c|}{BIC} & \multicolumn{4}{c}{AICc}\\ \hline
  & 1 & 2 & 3 & 4 & 1 & 2 & 3 & 4 \\ \hline
1 substructure  & 7 & 2 & 1 & 0 & 6 & 2 & 2 & 0 \\
2 substructures & 2 & 8 & 0 & 0 & 1 & 9 & 0 & 0 \\
3 substructures & 1 & 5 & 4 & 0 & 1 & 2 & 7 & 0 \\
4 substructures & 0 & 0 & 6 & 4 & 0 & 0 & 4 & 6 \\
\hline
\end{tabular}
\end{table}

\begin{table}
\centering
\caption{Summary of results for the Student-t and Student-t Masked simulations used to test efficacy of BIC and AICc in selecting the most appropriate model using our GMM analysis.  For each category, the number of simulated substructures in the mock catalog is listed down the left side, and the number of substructures in the preferred model is listed along the top.  In each cell of the table we list the number of simulations that yielded a preferred model with that number of substructures using BIC or AICc.}
\label{tab:GMMResults2}
\begin{tabular}{c|cccc|cccc}
\hline
\hline
\multicolumn{9}{c}{Student-t} \\ \hline
 & \multicolumn{4}{c|}{BIC} & \multicolumn{4}{c}{AICc}\\ \hline
 & 1 & 2 & 3 & 4 & 1 & 2 & 3 & 4 \\ \hline
1 substructure  & 9 & 0 & 1 & 0  & 4 & 2 & 3 & 1 \\
2 substructures & 0 & 10 & 0 & 0  & 0 & 7 & 3 & 0 \\
3 substructures & 0 & 4 & 6 & 0  & 0 & 3 & 5 & 2 \\
4 substructures & 1 & 2 & 3 & 4  & 0 & 0 & 4 & 4 \\
\hline
\hline
\multicolumn{9}{c}{Student-t Masked} \\ \hline
 & \multicolumn{4}{c|}{BIC} & \multicolumn{4}{c}{AICc}\\ \hline
 & 1 & 2 & 3 & 4 & 1 & 2 & 3 & 4 \\ \hline
1 substructure  & 90 & 19 & 1 & 0  & 55 & 45 & 8 & 2 \\
2 substructures & 9 & 99 & 2 & 0  & 3 & 91 & 13 & 3 \\
3 substructures & 5 & 60 & 45 & 0  & 1 & 44 & 61 & 4 \\
4 substructures & 4 & 49 & 41 & 16  & 2 & 44 & 46 & 18 \\
\hline
\end{tabular}
\end{table}

Let us first consider the results of the Gaussian simulations, where we expect our GMM analysis to perform well.  Use of the AICc produced more false positives than the BIC, with half (30\%) of the one substructure simulations being overestimated.  Conversely, using the AICc to choose the most preferred model does have a lower rate of underestimating the complexity of a system.  It is not surprising that underestimation becomes an issue when simulating three and four substructures, as the increase in the number of structures present means there is a higher incident of overlap, as well as a decrease in the number of galaxies ``sampled" for each substructure.

When we smooth the distribution of galaxies with our ``masking", the rate of overestimation is decreased slightly for both the BIC and AICc, while incidents of underestimation are increased.  This is again what is expected, as the smoothing of the galaxy distribution is more likely to blend structures together than separate them.  We did have a few incidents of single distributions being split by the ``masking" procedure.  In such cases, the additional substructure(s) were the result of splitting galaxies in RA-Dec projected space.  This was probably due to the ``masking" procedure creating enough of a void to separate a single distribution into multiple parts.  Because the ``masking" procedure did not utilize any velocity information, such voids were less likely to appear in the velocity distributions of the galaxies.  Therefore, our splitting of the Northern galaxies in ZwCl 2341.1+0000 is less likely to be an issue caused by our sampling of the system using DEIMOS slit masks, and more likely to be a proper separation of two substructures with a measurable line of sight velocity difference.

When we consider the Student-t simulations, we start to see a clearer difference between the BIC and AICc selected results.  Using the BIC, our GMM analysis was quite successful in obtaining the correct result in the one and two substructure simulations, with only a single incident of overestimation between them.  With the three and four substructure simulations, however, underestimation again became a prevalent issue, likely for the same reasons already discussed above.  In the AICc results, on the other hand, overestimation became a much larger issue, with 60\% of the the one substructure simulations being incorrectly split.  And while the AICc was again less likely to underestimate the complexity, that was a minor gain that was far outweighed by the overestimation rate.

The Student-t Masked simulations continue to show the clear difference when using the BIC versus the AICc in selecting preferred models.  With the BIC, our GMM analysis is quite successful at choosing the correct model for one and two substructure simulations.  The rate of underestimation, however, sharply increases with the three and four substructure simulations, likely due to the overlap, under-sampling, and smoothing issues already discussed.  Results using the AICc again have a much higher rate of overestimation for the simpler simulations, and a slightly lower rate of underestimation for the more complex ones.  This may appear to show that the AICc is more successful than the BIC at choosing correct models for more complex systems.  However, we also found that in approximately 25\% of cases when the BIC underestimated the number of substructures but the AICc selected a model with the correct number, at least one of the substructures in the more complex model was in the wrong location (i.e. one or more of the existing substructures were incorrectly split rather than correctly locating the missing substructure).

In an ideal scenario, our clustering analysis would be capable of identifying the correct substructures in a system every time.  Short of that, we find it is better to have a technique that always either underestimates or accurately locates the substructures present for two reasons.  Firstly, this allows us to set a firm lower limit on the complexity of the system.  In addition, to overestimate the number of substructures present in a system, a single distribution must be subdivided into multiple parts.  If the clustering analysis does this with any regularity, it reduces the confidence that the located substructures are actually real.  As we saw with a number of the tested Student-t Masked simulations, it is possible to choose a model with the correct number of substructures that is still wrong, because some have been artificially merged, while another has been subdivided.  Given that the BIC is less likely to overestimate the number of substructures, and is less likely to accept an incorrect model, we find it to be the superior criterion for use with our GMM analysis.

In all categories, overestimation stopped entirely once the simulations included four substructures.  This is probably due to the amount of overlap between the substructures when so many are crammed into such a small volume of space, combined with the relatively low number of galaxies sampled from each substructure.  It is possible that our method is only sensitive to up to four substructures in an equivalent volume of space with such a limited number of galaxies sampled.

\section{Over-splitting Bias in Dynamical Mass Estimates}
From the method of \cite{Girardi98}, the dynamical mass estimate of a cluster is calculated as:
\begin{equation}
M_{V} = \frac{3\pi}{2} \frac{\sigma^2_P R_{PV}}{G}
\end{equation}
where $\sigma_{P}$ and $R_{PV}$ are the line of sight velocity dispersion and projected virial radius of the cluster, respectively.  This equation is derived under the assumption that the cluster is spherically symmetric, has no anisotropy in the galaxy motions, and, most importantly, the galaxy orbits have virialized.  It is a well-recognized fact that these assumptions are violated in a merging galaxy cluster system because the galaxies in each cluster are subject to an outside gravitational force.  Because of this outside disturbance, a dynamical mass estimate for the system will likely be biased high.  In our study of ZwCl 2341.1+0000 we encountered another bias in dynamical mass estimates that we had not considered before.  This new bias comes from over-splitting a distribution of galaxies into too many smaller clusters, and will also lead to an over-estimation of a system's mass.

Examination of Eq. A1 can show why this occurs.  The dynamical mass estimate is calculated from two cluster properties, the velocity dispersion and the projected virial radius.  However, when choosing a subset of galaxies from a single cluster, only one of these values is likely to change.  The velocity dispersion for oversplit substructure made from a subset of galaxies is likely to be the same (or nearly so) as the velocity dispersion for the whole cluster.  The virial radius for each of these substructures, on the other hand, will change depending on how the galaxies are split, and should always be smaller than the virial radius of the original cluster.  Thus, the total mass estimate ($M_{V,tot}$) will increase as the sum of the virial radii, $R_{PV,i}$, of the substructures:

\begin{equation}
M_{V,tot} =\Sigma M_{V,i} = \frac{3\pi}{2} \frac{\sigma^2_P }{G} \sum R_{PV,i}
\end{equation}

If we divide this by the true mass of the cluster ($M_{V,true}$), then we obtain a ratio that is equal to the multiplicative factor, B, by which the true mass is over-estimated.

\begin{equation}
B = \frac{M_{V,tot}}{M_{V,true}} = \frac{\sum R_{PV,i}}{R_{PV,true}}
\end{equation}

The maximum value for this multiplicative factor is N, the number of substructures the cluster is being divided into.  It is unlikely for B to achieve this value, as it could only occur if $R_{PV,i} = R_{PV,true}$ for every substructure, and such a scenario calls for stacking N equally sized substructures directly on top of each other.  Similarly, the minimum value for B should be $\sqrt{N}$ because, if there is no overlap between the substructures, the sum of the areas covered by each substructure should be approximately equal to total area of the cluster.  For example, consider the simple scenario of splitting a single galaxy cluster into four equal quadrants.  In this case, the radius of each substructure would be half the radius of the whole distribution, and the total area covered by the four substructures would be equal to the area of the whole cluster.  Then, following our assumptions above, the total dynamical mass estimate would be double the true mass, which is consistent with $B = \sqrt{N}$.  However, it is unlikely that there will be no overlap between substructures in our Gaussian Mixture Model (GMM) analysis.  With this in mind, we expect B to be larger than $\sqrt{N}$.

To better characterize this over-splitting bias, we simulated a single galaxy cluster of known mass.  We then applied our GMM analysis to the simulated cluster to model it with a number of substructures, ranging from one to seven.  The total dynamical mass estimate was then calculated for each model.  This process was repeated over 10,000 iterations and the results were fit with a simple power law, shown below.

\begin{equation}
B = \alpha N^{\beta}
\end{equation}

Best fit values for this relation were $\alpha = 0.99526 \pm 0.00002$ and $\beta = 0.59154 \pm 0.000023$, which followed our expectations of $\alpha \sim 1$ and $\beta$ between 0.5 and 1.0.  

It is worth mentioning that these best fit values for the multiplicative factor are likely specific to our analysis process.  With our GMM analysis on a single simulated distribution of galaxies, it was very unlikely for the galaxies to be divided in velocity space.  If such a division did occur, like with the Northern galaxies of ZwCl 2341.1+0000, it would have a much larger effect on the total dynamical mass estimate than when we simply divide the galaxies in projected position space.  For such a scenario, we would not expect the over-splitting bias to behave the same as we have discussed above.  In fact, if a cluster were over-split in velocity space, it would be possible for the total mass estimate to be lower than the true mass of the cluster.  Similarly, if a different method is used to calculate the dynamical mass, the over-splitting bias may behave differently.  For example, from the method of \cite{Evrard08}, the dynamical mass is calculated as:

\begin{equation}
M_{200} = \left(\frac{\sigma_P}{\sigma_{DM,15}}\right)^{1/0.3361} \frac{10^{15}M_{\odot}}{h(z)}
\end{equation}
where everything but the projected velocity dispersion ($\sigma_P$) are constants.  Using this method, we would expect the total dynamical mass estimate for an over-split cluster to be N times the true mass, assuming that the projected velocity dispersion for each substructure remained the same as that of the whole cluster.

\bibliography{ZwCl2341_bib}

\label{lastpage}

\end{document}